\documentclass[fleqn,usenatbib,onecolumn]{rasti}

\usepackage{newtxtext,newtxmath}

\usepackage[T1]{fontenc}

\DeclareRobustCommand{\VAN}[3]{#2}
\let\VANthebibliography\thebibliography
\def\thebibliography{\DeclareRobustCommand{\VAN}[3]{##3}\VANthebibliography}

\usepackage{graphicx}	
\usepackage{amsmath}	
\usepackage{threeparttable}
\usepackage{booktabs}
\usepackage{makecell}
\usepackage{enumerate}
\usepackage[shortlabels]{enumitem}
\usepackage{color}
\usepackage{multirow}  
\usepackage[table]{xcolor}
\DeclareSymbolFont{CMletters}{OT1}{cmtt}{m}{ui}
\DeclareMathSymbol{\g}{\mathord}{CMletters}{`g}
\usepackage{caption}
\usepackage{bbding}
\usepackage{subcaption}
\usepackage{array}
\usepackage[section]{placeins}

\makeatother

\usepackage{listings}
\usepackage{xcolor}
\DeclareCaptionFormat{listing}{\textbf{#1}#2#3}
\definecolor{codegreen}{rgb}{0,0.6,0}
\definecolor{codegray}{rgb}{0.5,0.5,0.5}
\definecolor{codered}{rgb}{0.7,0.1,0.1}
\definecolor{codeblue}{rgb}{0,0,1}
\definecolor{codegreenblue}{rgb}{0.2,0.5,0.5}

\newcommand\pythonstyle{\lstset{
    language=Python,
    backgroundcolor=\color{white},   
    commentstyle=\color{codegreenblue}\itshape, 
    keywordstyle=\color{codegreen}\bfseries,    
    numberstyle=\tiny\color{codegray},
    stringstyle=\color{codered},
    basicstyle=\ttfamily,
    breakatwhitespace=false,         
    breaklines=false,                 
    captionpos=b,                    
    keepspaces=false,                 
    numbers=none,                    
    numbersep=5pt,                  
    showspaces=false,                
    showstringspaces=false,
    escapeinside={(*@}{@*)},
    showtabs=false,                  
    tabsize=2,
    emphstyle={\color{codeblue}},
    emph={[2]max, round, values, astype},
    emphstyle={[2]\color{codegreen}},
    emph={[3]numpy, np, ac},
    emphstyle={[3]\color{codeblue}\bfseries},
    emph={[4]as},
    emphstyle={[4]\color{codegreen}\bfseries},
    literate=
     {pyexocross}{{{\color{codeblue}\bfseries pyexocross}}}{8}
     {px}{{{\color{codeblue}\bfseries px}}}{1}
}
}

\lstnewenvironment{python}[1][]{
\pythonstyle \lstset{#1} }{}

\newcommand{\ExoMol}{{\textsc{ExoMol}}}
\newcommand{\ExoAtom}{{\textsc{ExoAtom}}}
\newcommand{\ExoMolHR}{{\textsc{ExoMolHR}}}

\newcommand{\PyExoCross}{{\textsc{PyExoCross}}}

\newcommand{\cm}{cm$^{-1}$}
\newcommand{\um}{$\mu$m}

\DeclareSymbolFont{CMletters}{OT1}{cmtt}{m}{ui}
\DeclareMathSymbol{\g}{\mathord}{CMletters}{`g}
\newcommand{\gf}{\g\!f}
\DeclareSymbolFont{matha}{OML}{txmi}{m}{it} 
\DeclareMathSymbol{\varv}{\mathord}{matha}{118}

\usepackage{pifont}
\newcommand{\cmark}{\ding{51}}
\newcommand{\xmark}{\ding{55}}

\usepackage{color}

\graphicspath{{./}{figures/}}

\title[\PyExoCross\ 2.0]{\PyExoCross\ 2.0: A Python Framework for LTE and non-LTE Spectra, Cross Sections, and Spectroscopic Post-processing of Atomic and Molecular Line Lists}

\author[Jingxin Zhang et al.]{
Jingxin Zhang,$^{1,2}$\thanks{E-mail: jingxin.zhang321@outlook.com (JZ)}
Sergei N. Yurchenko,$^{2}$
and Jonathan Tennyson$^{2}$\thanks{E-mail: j.tennyson@ucl.ac.uk (JT)}
\\
$^{1}$State Key Laboratory of High Temperature Gas Dynamics, Institute of Mechanics, Chinese Academy of Sciences, Beijing 100190, China\\
$^{2}$Department of Physics and Astronomy, University College London, Gower Street, WC1E 6BT London, UK
}

\date{Accepted XXX. Received YYY; in original form ZZZ}

\pubyear{\the\year{2026}}

\begin{document}
\label{firstpage}
\pagerange{\pageref{firstpage}--\pageref{lastpage}}
\maketitle

\begin{abstract}
\PyExoCross\ is a Python-based spectroscopic post-processing framework for converting large atomic and molecular line lists into scientifically useful quantities, including partition function, specific heats, cooling functions, lifetimes, oscillator strengths, line intensities, stick spectra, and absorption and emission cross sections. It is intended for applications in astrophysics, planetary atmospheres, laboratory spectroscopy, and other high-temperature environments where large modern databases require efficient and reproducible computational treatment.  
The new release provides both a configuration-file-driven command-line interface (CLI) and a Python application programming interface (API) within a unified computational framework. The original input file workflow is retained, while the Python API enables direct use in scripts, notebooks, and automated pipelines.
The version \PyExoCross\ 2.0 also adds explicit support for non-local thermodynamic equilibrium (non-LTE) calculations, including two-temperature models and user-defined density and population treatments for both absorption and emission spectra. In addition, database compatibility has been expanded beyond \ExoMol, HITRAN, and HITEMP line lists to include high-resolution molecular database \ExoMolHR\ and atomic database \ExoAtom. 
GPU acceleration is also introduced for computationally intensive intensity and cross-section calculations and is available for all currently supported database formats.
These developments improve workflow flexibility, reproducibility, and integration with modern data-analysis environments, while extending the physical modelling capabilities of the code. \PyExoCross\ therefore provides a more general and extensible platform for large-scale spectroscopic simulations based on modern atomic and molecular databases. 
\end{abstract}




\section{Introduction} \label{sec:intro}

The interpretation of atomic and molecular spectra is central to a wide range of applications in astrophysics, planetary science, laboratory spectroscopy, and high-temperature gas diagnostics. Modern spectroscopic databases such as \ExoMol \citep{jt528,jt939}, HITRAN \citep{jt1000}, HITEMP \citep{jt480}
and related resources now provide line lists of unprecedented size and scope, enabling detailed modelling of absorption and emission spectra over broad ranges of temperature, pressure, and wavelength. 
However, the practical use of these data requires efficient and reproducible post-processing tools capable of transforming large raw line lists into physically meaningful spectroscopic observables.

\PyExoCross\  was developed as a Python based framework for the post-processing of spectroscopic databases as an alternative to the Fortran code \textsc{ExoCross} \citep{jt708}. \PyExoCross\ supports  line list data format conversion and the calculation of partition function, specific heats, cooling functions, lifetimes, oscillator strengths, line intensities, stick spectra, and cross sections using a range of line profile models. The original published  version \citep{jt914} provided a convenient alternative to existing workflows for large scale spectral simulations. 
As database size, modelling demands, and workflow complexity have continued to increase, further development has become necessary.

This paper presents a new release of \PyExoCross\  which significantly extends the functionality of the code in several directions. 
First, in addition to the original command line program workflow, \PyExoCross\ now also supports direct use as a Python package, allowing integration into scripts, notebooks, pipelines, and automated analysis environments. 
Second, the new release introduces explicit support for non-local thermodynamic equilibrium (non-LTE) spectral calculations, including two-temperature and user-defined density and population models for both absorption and emission. 
Third, database support has been extended beyond standard \ExoMol, HITRAN, and HITEMP line lists to include molecular line lists provided by \ExoMolHR\ \citep{jt962,jtexomolhr2026} and atomic ones from \ExoAtom\ \citep{jt990}. Fourth,
use of GPUs is now available for computationally intensive calculations.
Together, these developments make \PyExoCross\ a more flexible and extensible framework for modern spectroscopic applications.

The aim of this paper is to describe the design, methodology, and capabilities of \PyExoCross, and to demonstrate how the new release improves workflow integration, physical modelling flexibility, and support for emerging spectroscopic data products.

\section{Methodology} \label{sec:software}

\subsection{Overall design} \label{sec:overall}

\PyExoCross\ is designed as a unified post-processing framework for large spectroscopic line lists and related state resolved data. The codebase extends the original input-file-driven workflow of \PyExoCross\ \citep{jt914} into an architecture that supports both a command-line interface (CLI) and a Python application programming interface (API).
Both interfaces use the same physical models, database parsers, computational routines, and output conventions.
The general design goal is to provide a single environment in which users can perform spectroscopic calculations from heterogeneous databases without developing database specific preprocessing procedures for each application.

The software is organized around three main layers. 
The first layer is a database interface layer, which parses source files from multiple spectroscopic databases and converts them into internal data structures with standardized field names and metadata descriptors. 
The second layer is a physics and spectroscopy layer, which implements the calculations of thermodynamic and radiative quantities, including partition functions, specific heats, cooling functions, lifetimes, oscillator strengths, stick spectra, and cross sections. As well as data format conversion between these various databases.
The third layer is a workflow layer, which exposes the common computational backend through both an input-file-driven command-line interface and a Python application programming interface. These interfaces are described in Section~\ref{sec:workflows}.

A major design principle of new version \PyExoCross\ is to isolate database specific syntax from spectroscopy specific computation. In practice, this implies that species formulas, quantum number labels and formats, and file naming conventions are normalized before entering the computation stage. This separation avoids duplication of functionality across supported databases and allows a similar computation backend to be reused for molecular and atomic data products. It also permits the incorporation of additional databases or updated file formats with minimal changes to the core computational algorithms.

Another design target is reproducible large scale batch processing. Spectroscopic calculations are frequently required over grids of temperatures, pressures, wavenumber or wavelength ranges, and line profile settings. 
\PyExoCross\ consequently emphasizes processes in which a single definition of the data source, species metadata, and computational parameters can be reused across multiple runs. 
For this reason, \PyExoCross\ also includes chunk based large file handling and automatic run logging. 
Large transition files can be streamed through the calculation pipeline with bounded memory use, and large outputs are written in chunks. 
The logging system records input settings, selected database and species, output formats, progress messages, timing information, and available CPU or GPU backend information. 
These features support reproducible long running calculations on workstations and high-performance computing systems.

Finally, \PyExoCross\ is designed to preserve backward compatibility while supporting incremental extension. The established input-file-driven workflow is retained, while fresh interfaces, database formats, and computational backends can be added without changing the underlying physical models. This architecture preserves existing user practices while expanding the software toward more automated, programmable, and multi-database spectroscopic analysis.

\subsection{Command-line and Python API workflows} \label{sec:workflows}

A major extension of \PyExoCross\ is the provision of two complementary user interfaces: an input-file-driven command-line interface (CLI) and a Python application programming interface (API). 
These are not separate implementations, but two access routes to the same computational backend. 

The original version of \PyExoCross\ was used through the input-file-driven CLI. This workflow is retained in the new release. It is suitable for standardized production calculations, shell scripts, repeated batch execution, and jobs on local workstations or high-performance computing (HPC) systems. Users specified the database, species, calculation type, physical parameters, and output settings in a configuration file, and then passed this file to the program from a terminal. 
For example,
\begin{verbatim}
$ python run.py -p input.inp
\end{verbatim}
The new Python API provides direct access to \PyExoCross\ functions after installation from PyPI: 
\begin{verbatim}
$ pip install pyexocross
\end{verbatim}
Users can then import the package in a Python script or notebook:
\begin{python}
import pyexocross as px
\end{python}
The imported package provides direct access to \PyExoCross\ functions without requiring a separate input file for each calculation.
Input data, intermediate representations, and calculated quantities can also be retained and manipulated as Python objects. The API is therefore useful for interactive analysis, parameter studies, customized workflows, and integration with data analysis and machine learning pipelines.

Both interfaces share the same routines for database parsing, state and transition handling, population and intensity calculations, line profile evaluation, and spectral synthesis. This common backend reduces code duplication, ensures methodological consistency between command-line and API calculations, and also allows users to switch workflows without changing the underlying physical models or output logic.
This shared architecture is particularly advantageous for extensive computations when file access and preprocessing contribute substantially to the total cost. 
When multiple outputs are required from the same line list, such as stick spectra and cross sections under identical physical conditions, the relevant files (which can be huge) can be read and processed once, allowing the intermediate representation to be reused in subsequent calculations. 
This reduces input/output operations and computational overhead in calculations spanning multiple temperatures, pressures, and spectral intervals.

Selected computationally intensive operations, including intensity and cross-section calculations, can also use GPU acceleration through either interface. The GPU implementation supports the database formats currently available in \PyExoCross\ and preserves the same physical definitions and output conventions as the CPU implementation.

\begin{figure}
\centering
\includegraphics[width=0.95\textwidth]{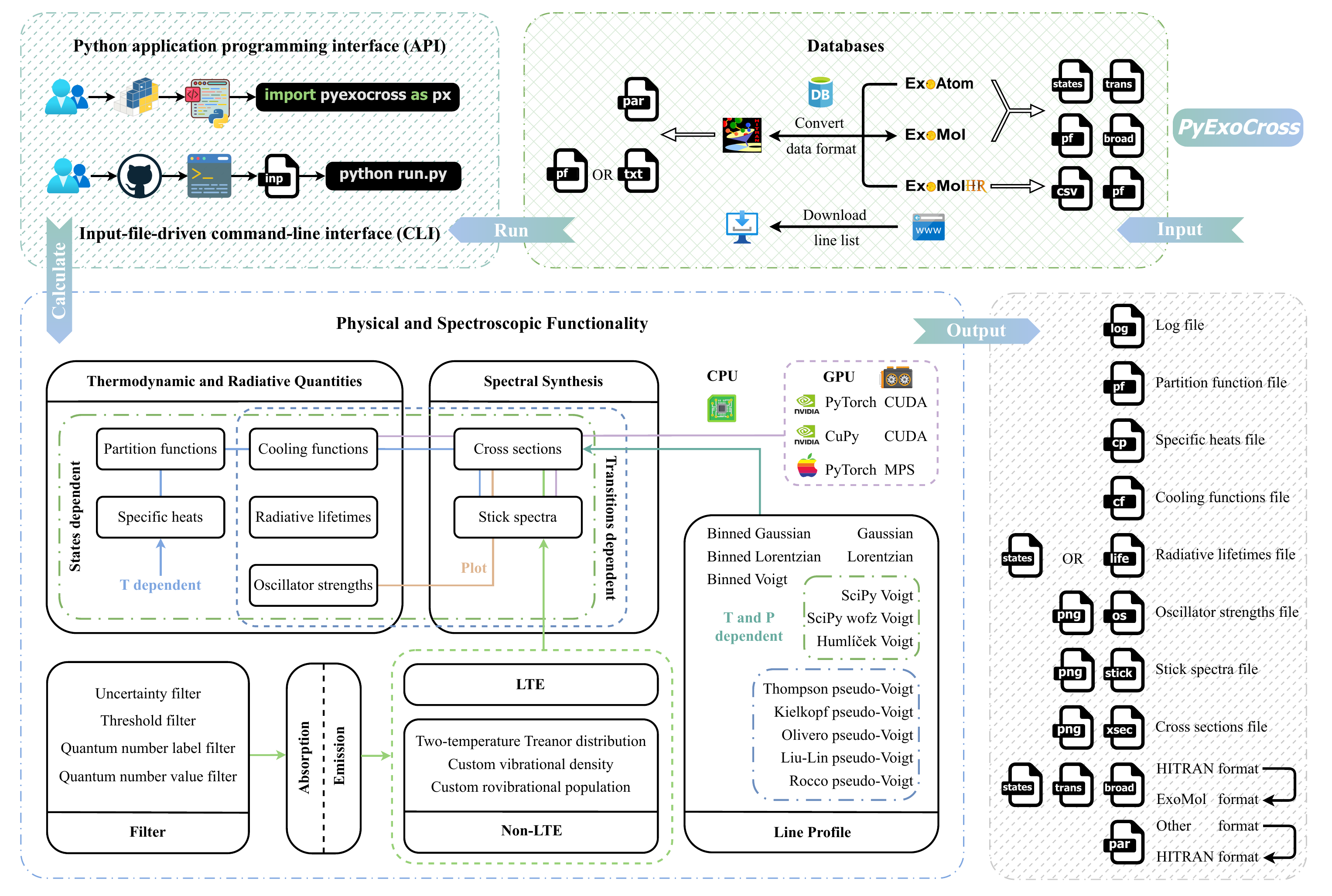}
\caption{\PyExoCross\ computational workflow for spectroscopic post-processing}
\label{fig:workflow}
\end{figure}

\PyExoCross\ is hosted on GitHub at \href{https://github.com/ExoMol/PyExoCross}{https://github.com/ExoMol/PyExoCross}. It is also distributed as a Python package via PyPI at \href{https://pypi.org/project/pyexocross}{https://pypi.org/project/pyexocross}, and documented online at \href{https://pyexocross.readthedocs.io}{https://pyexocross.readthedocs.io}.
The relationship between the two interfaces and the shared computational backend is summarized in Fig.~\ref{fig:workflow}. 
Reproducibility is supported by explicit parameter specification, standardized input and output conventions, and the use of a common backend across both interfaces.

\subsection{Supported databases and unified data interfaces} \label{sec:database}

\PyExoCross\ supports several spectroscopic databases with different file organizations, naming conventions, and metadata models. In the present version, these include the main \ExoMol\ molecular database, the HITRAN and HITEMP molecular databases, the \ExoMolHR\ high-resolution molecular database, and the \ExoAtom\ atomic database. Although these resources differ substantially in data layout and intended use cases, they are handled within \PyExoCross\ through a unified parsing and metadata mapping strategy.

The role of the database interface is not only to read files, but also to map heterogeneous source data to a common internal representation suitable for subsequent calculations. This includes standardization of species identifiers, quantum numbers, statistical weights, energy levels, Einstein $A$-coefficients, and auxiliary metadata required by downstream modules. The database specific differences are therefore localized within dedicated import, export, and mapping routines, whereas the higher level physical calculations operate on normalized internal structures.

This section summarizes how each supported database is represented within \PyExoCross\ and how these representations are reconciled through a common interface. 
Throughout this section, angle-bracketed terms denote placeholders used in database file names. 
\texttt{<ISO-SLUG>}, \texttt{<ATOM-SLUG>}, and \texttt{<DATASET>} denote the molecular isotopologue identifier, atomic species identifier, and dataset identifier, respectively.

\subsubsection{\ExoMol\ molecular database} \label{sec:exomol}

\ExoMol \footnote{\ExoMol\ database website: \href{https://www.exomol.com}{https://www.exomol.com}} database provides large scale theoretical and empirically refined molecular line lists designed primarily for hot atmospheres and other high-temperature applications \citep{jt528}. Its standard data structure \citep{jt548,jt1032} separates molecular information into states files (\texttt{.states}), transitions files (\texttt{.trans}), partition function files (\texttt{.pf}), broadening files (\texttt{.broad}), and definition files (\texttt{.def.json}). This organization is particularly suitable for very large line lists, since the transitions data can be distributed over multiple files separated by wavenumber ranges while sharing a common set of state energy levels and associated metadata such as quantum numbers.  
Table~\ref{tab:exomolstates} and Table~\ref{tab:exomoltrans} show the data formats of \ExoMol\ line list states and transitions files. Table~\ref{tab:exomolbroad} specifies the \ExoMol\ broadening file format.

\begin{table}
\centering
\caption{Specification of the \ExoMol\ states file (\texttt{<ISO-SLUG>\_\_<DATASET>.states.bz2} 
including extra data options; the formats at the end of the table are for the compulsory section only \citep{jt939}.}
\label{tab:exomolstates}
\begin{threeparttable}
\setlength{\tabcolsep}{5.3mm}
\begin{tabular}{llll}
\toprule
Field & Fortran Format & C Format & Description \\
\midrule
ID                & \texttt{I12}           & \texttt{\%12d}         & State ID \\
$\tilde{E}$       & \texttt{F12.6}         & \texttt{\%12.6f}       & Recommended state energy in \cm \\
$\g_\mathrm{tot}$ & \texttt{I6}            & \texttt{\%6d}          & Total state degeneracy \\
$J$               & \texttt{I7/F7.1}       & \texttt{\%7d/\%7.1f}   & Total angular momentum quantum number, $J$ or $F$ (integer/half-integer ) \\
$\Delta E$        & \texttt{F12.6}         & \texttt{\%12.6f}       & Uncertainty in the state energy in \cm \\
$\tau$            & \texttt{ES12.4}        & \texttt{\%12.4E}       & State lifetime (aggregated radiative and predissociative lifetimes) in s \\
($\g$)            & \texttt{F10.6}         & \texttt{\%10.6f}       & Land\'e $\g$-factor (optional) \\
(QN)              & See definition file    & See definition file    & State quantum numbers, may be several columns (optional) \\
(Abbr)            & \texttt{A2}            & \texttt{\%2s}          & Abbreviation giving source of state energy \citep{jt948} (optional) \\
($\tilde{E}_\textrm{cal}$) & \texttt{F12.6}         & \texttt{\%12.6f}       & Calculated state energy in \cm\ (optional) \\
\bottomrule
\end{tabular}
\begin{tablenotes}[flushleft]
\footnotesize
\item Quantum number labels and formats as defined in the \ExoMol\ definition file, \texttt{<ISO-SLUG>\_\_<DATASET>.def.json}. 
\item Fortran format for text file: $J$ integer: \texttt{(I12,1x,F12.6,1x,I6,I7,1x,F12.6,1x,ES12.4,1x,F10.6)} 
\item or $J$ helf-integer: \texttt{(I12,1x,F12.6,1x,I6,F7.1,1x,F12.6,1x,ES12.4,1x,F10.6)}. 
\end{tablenotes}
\end{threeparttable}
\end{table}

\begin{table}
\centering
\caption{Specification of the \ExoMol\ transitions file (\texttt{<ISO-SLUG>\_\_<DATASET>*.trans.bz2}) including extra data options \citep{jt939}.}
\label{tab:exomoltrans}
\begin{threeparttable}
\setlength{\tabcolsep}{5.7mm}
\begin{tabular}{llll}
\toprule
Field & Fortran Format & C Format & Description \\
\midrule
$u$ & \texttt{I12} & \texttt{\%12d} & Upper state ID \\
$l$ & \texttt{I12} & \texttt{\%12d} & Lower state ID \\
$A$ & \texttt{ES10.4} & \texttt{\%10.4E} & Einstein $A$-coefficient in $\mathrm{s^{-1}}$ \\
$\tilde{\nu}_{ul}$ & \texttt{ES15.6} & \texttt{\%15.6E} & Transition wavenumber in \cm\  (optional) \\
\bottomrule
\end{tabular}
\begin{tablenotes}[flushleft]
\footnotesize
\item Transitions filenames are usually in format \texttt{<ISO-SLUG>\_\_<DATASET>.trans.bz2}. However, for large line lists, the filenames use the wavenumber range to split the whole transitions into smaller segments, and they are in format  \texttt{<ISO-SLUG>\_\_<DATASET>\_\_<min}$\nu$\texttt{>-<max}$\nu$\texttt{>.trans.bz2}. 
\item Fortran format for text file: \texttt{(I12,1x,I12,1x,ES10.4,1x,ES15.6)}. 
\end{tablenotes}
\end{threeparttable}
\end{table}

\begin{table}
\centering
\caption{Specification of the mandatory part of the \ExoMol\ broadening file (\texttt{<ISO-SLUG>\_\_*.broad}) format \citep{jt939}.}
\label{tab:exomolbroad}
\begin{threeparttable}
\setlength{\tabcolsep}{7.5mm}
\begin{tabular}{llll}
\toprule
Field & Fortran Format & C Format & Description \\
\midrule
code                    & \texttt{A2}      & \texttt{\%2s}        & Code identifying quantum number set following $J$ \\
$\gamma_{\textrm{ref}}$ & \texttt{F6.4}    & \texttt{\%6.4f}      & Lorentzian half-width at reference temperature and pressure in \cm$/\textrm{bar}$ \\
$n_{\textrm{L}}$        & \texttt{F6.3}    & \texttt{\%6.3f}      & Temperature exponent \\
$J''$                   & \texttt{I7/F7.1} & \texttt{\%7d/\%7.1f} &  Lower $J$-quantum number integer/half-integer \\
\bottomrule
\end{tabular}
\begin{tablenotes}[flushleft]
\footnotesize
\item * in broadending files can be \texttt{air}, \texttt{self}, some molecules like \texttt{H2} and \texttt{CO2}, or some atoms like \texttt{Ar} and \texttt{He}. 
\item Fortran format for text file: $J$ integer: \texttt{(A2,1x,F6.4,1x,F6.3,1x,I7)} or $J$ half-integer: \texttt{(A2,1x,F6.4,1x,F6.3,1x,F7.1)}. 
\end{tablenotes}
\end{threeparttable}
\end{table}

\subsubsection{HITRAN and HITEMP molecular databases} \label{sec:hitran}

HITRAN \footnote{HITRAN database website: \href{https://hitran.org}{https://hitran.org}} and HITEMP \footnote{HITEMP database website: \href{https://hitran.org/hitemp}{https://hitran.org/hitemp}} databases are widely used molecular spectroscopic databases with formats and metadata conventions which differ from those of \ExoMol. In contrast to the \ExoMol\ organization, HITRAN and HITEMP line lists are traditionally distributed in a compacted line-by-line format in the form of \texttt{.par} file in which each transition record contains the parameters required for direct spectroscopic use.
Table~\ref{tab:hitranpar} gives an overview of the standard format; other, self-designed formats are possible \citep{HAPI}. These databases are especially important for terrestrial, laboratory, and high-temperature applications in which pressure broadening parameters, reference intensities, and standardized labels are part of the source data model.

\begin{table}
\centering
\caption{Specification of the HITRAN/HITEMP line-by-line format text file (\texttt{*.par}) in 2004 Edition \citep{hitran2004} (160-character record).}
\label{tab:hitranpar}
\begin{threeparttable}
\setlength{\tabcolsep}{7.6mm}
\begin{tabular}{llll}
\toprule
Field & Fortran Format & C Format & Description \\
\midrule
M & \texttt{I2} & \texttt{\%2d} & Molecule number, HITRAN chronological assignment \\
I & \texttt{I1} & \texttt{\%1d} & Isotopologue number, ordering within a molecule by terrestrial abundance \\
$\nu$ & \texttt{F12.6} & \texttt{\%12.6f} & Vacuum wavenumber in \cm \\
$S$ & \texttt{ES10.3} & \texttt{\%10.3e} & Intensity in \cm/(molecule cm$^{-2}$) at standard $296$ K \\
$A$ & \texttt{ES10.3} & \texttt{\%10.3e} & Einstein $A$-coefficien in s$^{-1}$ \\
$\gamma_{\textrm{air}}$ & \texttt{F5.4} & \texttt{\%5.4f} & Air-broadened half-width (HWHM) in \cm atm$^{-1}$ at $296$ K \\
$\gamma_{\textrm{self}}$ & \texttt{F5.4} & \texttt{\%5.4f} & Self-broadened half-width (HWHM) in \cm atm$^{-1}$ at $296$ K \\
$E''$ & \texttt{F10.4} & \texttt{\%10.4f} & Lower-state energy in \cm \\
$n_{\textrm{air}}$ & \texttt{F4.2} & \texttt{\%4.2f} & Temperature-dependent exponent for $\gamma_{\textrm{air}}$ \\
$\delta_{\textrm{air}}$ & \texttt{F8.6} & \texttt{\%8.6f} & Air pressure-induced line shift in \cm atm$^{-1}$ at $296$ K \\
$V'$ & \texttt{A15} & \texttt{\%15s} & Upper-state global `quanta' \\
$V''$ & \texttt{A15} & \texttt{\%15s} & Lower-state global `quanta' \\
$Q'$ & \texttt{A15} & \texttt{\%15s} & Upper-state local `quanta' \\
$Q''$ & \texttt{A15} & \texttt{\%15s} & Lower-state local `quanta' \\
I$_{\textrm{err}}$ & \texttt{I6} & \texttt{\%6d} & Uncertainty indices, accuracy for 6 critical parameters \\
I$_{\textrm{ref}}$ & \texttt{I12} & \texttt{\%12d} & Reference indices, references for 3/6 critical parameters \\
$*$ & \texttt{A1} & \texttt{\%1s} & Character, availability of program and data for the case of line mixing \\
$\g'$ & \texttt{F7.1} & \texttt{\%7.1f} & The statistical weight of the upper state \\
$\g''$ & \texttt{F7.1} & \texttt{\%7.1f} & The statistical weight of the lower state \\
\bottomrule
\end{tabular}
\begin{tablenotes}[flushleft]
\footnotesize
\item *: The filename is usually named with random characters. 
\item Fortran format: \texttt{(I2,I1,F12.6,ES10.3,ES10.3,F5.4,F5.4,F10.4,F4.2,F8.6,A15,A15,A15,A15,I6,I12,A1,F7.1,F7.1)}. 
\end{tablenotes}
\end{threeparttable}
\end{table}

\subsubsection{\ExoMolHR\ accurate high-resolution molecular database} \label{sec:exomolhr}

\ExoMolHR \footnote{\ExoMolHR\ database website: \href{https://www.exomol.com/exomolhr}{https://www.exomol.com/exomolhr}} \citep{jt962,jtexomolhr2026}
is a high-resolution molecular database designed to provide subsets of transitions with improved line-position accuracy for applications in which spectroscopic precision is critical. In contrast to the \ExoMol\ structure, which is optimized for comprehensive line list storage, \ExoMolHR\ provides processed high-resolution products intended for efficient use in line identification, precision simulation, and related analyses. 
Table~\ref{tab:exomolhrcsv} specifies the \ExoMolHR\ line list file format.

From the perspective of \PyExoCross\ architecture, \ExoMolHR\ support demonstrates the usefulness of separating data ingestion from physical calculation. Once the high-resolution transition records and their metadata are mapped into the common internal representation, the same downstream machinery used for other databases can be applied. This permits \ExoMolHR\ data to participate in the same package workflows, scripted analyses, and output pipelines as standard molecular line lists, while preserving the distinctive role of the database as a precision oriented source.

\begin{table}
\centering
\caption{Specification of the \ExoMolHR\ line list file (precomputed \texttt{<MOLECULE>\_\_<ISO-SLUG>\_\_<DATASET>.csv/.parquet} or calculated \texttt{<YYYYMMDDHHMMSS>\_\_<ISO-SLUG>\_\_<T>K\_\_wn/wl<MIN\_WN/WL>-<MAX\_WN/WL>\_\_Smin<MIN\_S>.csv}) format \citep{jt962,jtexomolhr2026}.}
\label{tab:exomolhrcsv}
\begin{threeparttable}
\setlength{\tabcolsep}{6.6mm}
\begin{tabular}{llll}
\toprule
Field & Fortran Format & C Format & Description \\
\midrule
$\tilde{\nu}$ & \texttt{F12.6} & \texttt{\%12.6f} & Transition wavenumber in \cm \\
$S$    & \texttt{ES10.4}  & \texttt{\%10.4E}     &  Absorption intensity at the user specified temperature in cm$/$molecule \\
$\Delta \nu$ & \texttt{F12.6} & \texttt{\%12.6f} & Unceratinty in the transition wavenumber in \cm \\
$R$    & \texttt{F14.2}   & \texttt{\%14.2f}     & Resolving power $R>100\,000$ \\
$A$    & \texttt{ES10.4}  & \texttt{\%10.4E}     & Einstein $A$-coefficient in s$^{-1}$ \\
$E''$  & \texttt{F12.6}   & \texttt{\%12.6f}     & Lower state energy in cm$^{-1}$ \\
$\g'$   & \texttt{I6}     & \texttt{\%6d}        & Total state degeneracy for upper state \\
$\g''$  & \texttt{I6}     & \texttt{\%6d}        & Total state degeneracy for lower state \\
$J'$   & \texttt{I7/F7.1} & \texttt{\%7d/\%7.1f} & Total angular momentum quantum number (integer/half-integer) for upper state \\
$J''$  & \texttt{I7/F7.1} & \texttt{\%7d/\%7.1f} & Total angular momentum quantum number (integer/half-integer) for lower state \\
QN$'$  & *                & *                   & Quantum number for upper state \\
QN$''$ & *                & *                   & Quantum number for lower state \\
\bottomrule
\end{tabular}
\begin{tablenotes}[flushleft]
\footnotesize
\item *: Quantum number labels, formats, and descriptions are defined in tables on \ExoMolHR\ website quantum number information page \href{https://www.exomol.com/exomolhr/qn}{https://www.exomol.com/exomolhr/qn}. 
\item \texttt{<YYYYMMDDHHMMSS>} means year, month, date, hour, minute, and second, which is the created time of files. 
\item \texttt{T} is the user specified temperature which is an integer. 
\end{tablenotes}
\end{threeparttable}
\end{table}

\subsubsection{\ExoAtom\ atomic database} \label{sec:exoatom}

\ExoAtom \footnote{\ExoAtom\ database website: \href{https://www.exomol.com/exoatom}{https://www.exomol.com/exoatom}} database extends the database coverage of \PyExoCross\ from molecular to atomic spectroscopy. Atomic datasets differ from molecular line lists not only in their physical structure, but also in the conventions used to describe levels, transitions, and associated quantum identifiers. Supporting atomic data within the same framework therefore requires a metadata layer that is flexible enough to represent both molecular and atomic cases without forcing them into an overly restrictive common syntax. 
Table~\ref{tab:exoatomstates} and Table~\ref{tab:exoatomtrans} specify the data format of \ExoAtom\ line list states and transitions files.

In \PyExoCross, \ExoAtom\ data are incorporated through the same general database interface philosophy used for molecular resources. The parser identifies the relevant state and transition descriptors, standardizes the core radiative quantities, and exposes them to the shared calculation backend. This allows atomic spectra and related quantities to be generated within the same software environment as molecular spectra, enabling consistent workflows across different classes of spectroscopic species.

The inclusion of \ExoAtom\ is also methodologically important because it broadens the scope of the software from a molecular line list post-processor to a more general spectroscopic post-processing platform. This is particularly useful for applications in which molecular and atomic signatures must be treated within a single computational workflow.

\begin{table}
\centering
\caption{Specification of the \ExoAtom\ states file (\texttt{<ATOM-SLUG>\_\_<DATASET>.states}) from NIST \citep{NIST} and Kurucz \citep{11Kurucz.db} sources including extra data options; the formats at the end of the table are for the compulsory section only \citep{jt990}.}
\label{tab:exoatomstates}
\begin{threeparttable}
\setlength{\tabcolsep}{6mm}
\begin{tabular}{llll}
\toprule
Field & Fortran Format & C Format & Description \\
\midrule
ID               & \texttt{I12}         & \texttt{\%12d}           & State ID \\
$\tilde{E}$      & \texttt{F12.6}       & \texttt{\%12.6f}         & State energy in \cm \\
                 & \texttt{F12.5/F12.4} & \texttt{\%12.5f/\%12.4f} & \\
$\g_J$           & \texttt{I6}          & \texttt{\%6d}            & State degeneracy \\
$J$              & \texttt{I7/F7.1}     & \texttt{\%7d/\%7.1f}     & Total angular momentum quantum number, $J$ (integer/half-integer) \\
$\Delta E$       & \texttt{F12.6}       & \texttt{\%12.6f}         & Uncertainty in the state energy in \cm \\
($\tau$)         & \texttt{ES12.4}      & \texttt{\%12.4e}         & Radiative lifetime in s (optional, Kurucz only) \\
(gfactor)        & \texttt{F10.6}       & \texttt{\%10.6f}         & Land\'e $\g$-factor (optional, Kurucz only) \\
qn:configuration & \texttt{A12}         & \texttt{\%12s}           & Configuration for the state \\
term             & \texttt{A8}          & \texttt{\%8s}            & Term for the state \\
(qn:parity)      & \texttt{A1}          & \texttt{\%1s}            & Parity for the state (optional, NIST only) \\
(Abbr)           & \texttt{A2}          & \texttt{\%2s}            & Abbreviation indicating data source: \\
                 &                      &                          & \texttt{CA} (calculated, Kurucz) or \texttt{NI} (measured, NIST) \\
\bottomrule
\end{tabular}
\begin{tablenotes}[flushleft]
\footnotesize
\item \ExoAtom\ database has two data sources which are NIST and Kurucz databases. 
\item The exact details and file formats can be found from corresponding definition \texttt{.adef.json} files.
\item $\tilde{E}$: Different formats apply: $\tilde{E} \leq 100000$: \texttt{F12.6 or \%12.6f}, $100000 \leq \tilde{E} < 1000000$: \texttt{F12.5 or \%12.5f}, $\tilde{E} \geq 1000000$: \texttt{F12.4 or \%12.4f}. 
\item $\g_J$: Obtained directly from $\g$ in levels data. When the atom has isotopes, it corresponds to $\g_{\textrm{tot}}$; when the atom does not have isotopes, it corresponds to $\g_J$. 
\item term: Directly from the term column in levels data. The trailing `*' should be removed if applicable. 
\item qn:parity: Terms of odd parity (those ending in * in the original data sources) are marked with \texttt{-} in this field; those of even parity are marked with \texttt{+}. 
\item Abbr: Only in Kurucz-based files; indicates whether the level is experimentally identified (\texttt{NI}) or purely calculated (\texttt{CA}). 
\item Fortran format for text file: $J$ integer: \texttt{(I12,1x,F12.6/F12.5/F12.4,1x,I6,1x,I7,1x,F12.6,1x,ES12.4,1x,F10.6,1x,A12,1x,A8,1x,A1,1x,A2)} 
\item or $J$ half-integer: \texttt{(I12,1x,F12.6/F12.5/F12.4,1x,I6,1x,F7.1,1x,F12.6,1x,ES12.4,1x,F10.6,1x,A12,1x,A8,1x,A1,1x,A2)}.
\end{tablenotes}
\end{threeparttable}
\end{table}

\begin{table}
\centering
\caption{Specification of the \ExoAtom\ transitions file (\texttt{<ATOM-SLUG>\_\_<DATASET>.trans}) \citep{jt990}.}
\label{tab:exoatomtrans}
\begin{threeparttable}
\setlength{\tabcolsep}{7.2mm}
\begin{tabular}{llll}
\toprule
Field & Fortran Format & C Format & Description \\
\midrule
$u$ & \texttt{I12} & \texttt{\%12d} & Upper state ID \\
$l$ & \texttt{I12} & \texttt{\%12d} & Lower state ID \\
$A$ & \texttt{ES10.4} & \texttt{\%10.4E} & Einstein $A$ coefficient in $\mathrm{s^{-1}}$ \\
$\tilde{\nu}_{ul}$ & \texttt{ES15.6} & \texttt{\%15.6E} & Transition wavenumber in \cm. \\
\bottomrule
\end{tabular}
\begin{tablenotes}[flushleft]
\footnotesize
\item Fortran format for text file: \texttt{(I12,1x,I12,1x,ES10.4,1x,ES15.6)}. 
\end{tablenotes}
\end{threeparttable}
\end{table}

\subsubsection{Unified species identification and metadata mapping} \label{sec:meta}

Because the supported databases adopt different naming rules, metadata fields, and quantum number conventions, a unified species-identification layer is required in order to expose a consistent user-facing workflow. In \PyExoCross, this is handled by metadata-mapping routines that translate database-native labels into standardized internal descriptors. These descriptors are then used throughout the rest of the program, irrespective of the original source database.

This unified metadata layer serves two practical purposes. From the perspective of users, it reduces the need to remember database-specific syntax when switching between resources. From the software perspective, it isolates database-specific heterogeneity from the calculation layer and therefore improves maintainability, extensibility, and reproducibility. In other words, the common internal metadata model is one of the key mechanisms that allows \PyExoCross\ to function as a unified multi-database spectroscopic framework rather than as a collection of independent file readers.

\subsection{Physical and spectroscopic functionality} \label{sec:functionality}

The physical and spectroscopic modules of \PyExoCross\ operate on the normalized data structures produced by the database interface layer. These modules implement a set of commonly required post-processing tasks for line list based spectroscopy, ranging from thermodynamic quantities to radiative diagnostics and simulated spectra. The emphasis is on providing a consistent computational environment in which the same species and metadata definitions can be reused across multiple types of calculations.

In the present work, the functionality implemented includes data-format conversion, partition functions, specific heats, cooling functions, lifetimes, oscillator strengths, stick spectra, and cross sections. 
Table~\ref{tab:file type} gives all related file types and Table~\ref{tab:file fmt} summarizes their corresponding output data format to the \PyExoCross. 
The corresponding definitions are introduced in the following sections, while the notation follows Table~\ref{tab:notation}.
Although these quantities have different physical meanings and numerical requirements, they are linked operationally by the same database interface and workflow infrastructure. This avoids repeated preprocessing of the source files and makes it possible to embed different calculations within a common program or package workflow.

\begin{table}
\centering
\caption{Specification of the input and output file types related to the \PyExoCross\ (contents in brackets are optional).}
\label{tab:file type}
\begin{threeparttable}
\setlength{\tabcolsep}{2.15mm}
\begin{tabular}{llcll}
\toprule
File extension & \PyExoCross\ available & $N_{\textrm{files}}$ & File Types & Contents \\
\midrule
\multicolumn{5}{c}{\textbf{\ExoMol\ file format: \ExoMol}} \\
\texttt{.def.json}   & Input        & 1    & Definition          & Defines contents of other files for each isotopologue \\
\texttt{.states.bz2} & Input/Output & 1    & States              & Energy level, quantum number, lifetime, (uncertainty, Land\'e $\g$-factor) \\
\texttt{.trans.bz2}  & Input/Output & $^a$ & Transitions         & Einstein $A$-coefficient, (wavenumber) \\
\texttt{.pf}         & Input/Output & 1    & Partition function  & Temperature-dependent partition function, (cooling function) \\
\midrule
\multicolumn{5}{c}{\textbf{\ExoAtom\ file format: \ExoAtom}} \\
\texttt{.adef.json} & Input        & 1    & Definition          & Defines contents of other files for each isotopologue \\
\texttt{.states}    & Input/output & 1    & States              & Energy level, quantum number, uncertainty, lifetime, (Land\'e $\g$-factor) \\
\texttt{.trans}     & Input        & $^a$ & Transitions         & wavenumber, Einstein $A$-coefficient \\
\texttt{.pf}        & Input/Output & 1    & Partition function  & Temperature-dependent partition function, discontinuous in Kurucz dataset \\
\midrule
\multicolumn{5}{c}{\textbf{\ExoMolHR\ file format: \ExoMolHR}} \\
\texttt{.csv/.parquet} & Input     & 1    & Line list file      & Accurate and high-resolution line list with intensity and resolving power \\
\texttt{.pf}        & Input/Output & 1    & Partition function  & Temperature-dependent partition function same as \ExoMol \\
\midrule
\multicolumn{5}{c}{\textbf{HITRAN file format: HITRAN and HITEMP}} \\
\texttt{.par}       & Input/Output & 1    & Line list file      & HITRAN line list file with intensity at $296$ K \\
\texttt{.pf}        & Input/Output & 1    & Partition function  & Temperature-dependent partition function, \texttt{.txt} can also be input file type \\
\midrule
\multicolumn{5}{c}{\textbf{Global file format}} \\
\texttt{.broad}      & Input/Output & $^b$ & Broadening          & Parameters for pressure-dependent line profiles \\
\texttt{.cp}         & Output       & 1    & Specific heat       & Temperature-dependent specific heat \\
\texttt{.cf}         & Output       & 1    & Cooling function    & Temperature-dependent cooling function \\
\texttt{.os}         & Output       & 1    & Oscillator strength & Oscillator strength, wavenumber \\
\texttt{.stick}      & Output       & $^c$ & Stick spectra       & Wavenumber or wavelength, intensity, energy level, quantum number \\
\texttt{.xsec}       & Output       & $^d$ & Cross sections      & Wavenumber or wavelength, cross section \\
\bottomrule
\end{tabular}
\begin{tablenotes}[flushleft]
\footnotesize
\item $N_{\textrm{files}}$ total number of possible files for each species. 
\item $^a$ There is a single \texttt{.trans} file, but for molecules with large numbers of transitions, it is subdivided by wavenumber regions. 
\item $^b$ There are $N_{\textrm{broad}}$ sets of \texttt{.broad} files for each species. 
\item $^c$ There are $N_{\textrm{stick}}$ sets of \texttt{.stick} files for each species, the number depends on the number of temperatures. 
\item $^d$ There are $N_{\textrm{xsec}}$ sets of \texttt{.xsec} files for each species, the number depends on the number of temperatures (and pressures). 
\end{tablenotes}
\end{threeparttable}
\end{table}
\begin{table}
\centering
\caption{Specification of the file formats for the output partition functions (\texttt{.pf}), specific heats (\texttt{.cp}), cooling functions (\texttt{.cf}), lifetimes (\texttt{.states}), oscillator strengths (\texttt{.os}), stick spectra (\texttt{.stick}), and cross sections (\texttt{.xsec}).}
\label{tab:file fmt}
\begin{threeparttable}
\setlength{\tabcolsep}{4.1mm}
\begin{tabular}{llll}
\toprule
Field & Fortran Format & C Format & Description \\
\midrule
\multicolumn{4}{c}{\textbf{Partition functions \texttt{.pf}}} \\
\multicolumn{4}{c}{Fortran format: (\texttt{F8.1,1x,F15.4})} \\
\midrule
$T$    & \texttt{F8.1}  & \texttt{\%8.1f}  & Temperature in K \\
$Q(T)$ & \texttt{F15.4} & \texttt{\%15.4f} & Partition function (dimensionless) \\
\midrule
\multicolumn{4}{c}{\textbf{Specific heats \texttt{.cp}}} \\
\multicolumn{4}{c}{Fortran format: (\texttt{F8.1,1x,F15.4})} \\
\midrule
$T$    & \texttt{F8.1}  & \texttt{\%8.1f}  & Temperature in K \\
$C_p(T)$ & \texttt{F15.4} & \texttt{\%15.4f} & Specific heat (dimensionless) \\
\midrule
\multicolumn{4}{c}{\textbf{Cooling functions \texttt{.cf}}} \\
\multicolumn{4}{c}{Fortran format: (\texttt{F8.1,1x,ES20.8})} \\
\midrule
$T$    & \texttt{F8.1}  & \texttt{\%8.1f}  & Temperature in K \\
$W(T)$ & \texttt{ES20.8} & \texttt{\%20.8E} & Cooling function in erg (s molecule sr)$^{-1}$ \\
\midrule
\multicolumn{4}{c}{\textbf{Lifetimes \texttt{.states} or \texttt{.states.bz2}}} \\
\multicolumn{4}{c}{Same as input states file} \\
\midrule
ID               & \texttt{I12}     & \texttt{\%12d}       & State ID \\
$\tilde{E}$      & \texttt{F12.6}   & \texttt{\%12.6f}     & Recommended state energy in \cm \\
$\g_\mathrm{tot}$ & \texttt{I6}      & \texttt{\%6d}        & Total state degeneracy \\
$J$              & \texttt{I7/F7.1} & \texttt{\%7d/\%7.1f} & Total angular momentum quantum number, $J$ or $F$ (integer/half-integer) \\
$\Delta E$       & \texttt{F12.6}   & \texttt{\%12.6f}     & Uncertainty in the state energy in \cm \\
$\tau$           & \texttt{ES12.4}  & \texttt{\%12.4E}     & State lifetime (aggregated radiative and predissociative lifetimes) in s \\
*                & *                & *                    & All same as the input states file \\
\midrule
\multicolumn{4}{c}{\textbf{Oscillator strengths \texttt{.os} and plots \texttt{.png}}} \\
\multicolumn{4}{c}{Fortran format: (\texttt{I12,1x,I12,1x,ES12.4,1x,F15.6})} \\
\midrule
$u$ & \texttt{I12} & \texttt{\%12d} & Upper state degeneracy \\
$l$ & \texttt{I12} & \texttt{\%12d} & Lower state degeneracy \\
$\gf$ or $f$ & \texttt{ES12.4} & \texttt{\%12.4E} & Weighted or actual oscillator strength \\
$\tilde{\nu}$ & \texttt{F15.6} & \texttt{\%15.6f} & Central bin wavenumber in \cm \\
\midrule
\multicolumn{4}{c}{\textbf{Stick spectra \texttt{.stick} and plots \texttt{.png}}} \\
\multicolumn{4}{c}{Fortran format: (\texttt{F15.6/ES15.8,1x,ES15.8,1x,I7/F7.1,1x,F12.6,1x,I7/F7.1,1x,F12.6...})} \\
\midrule
$\tilde{\nu}$ or $\tilde{\lambda}$ & \texttt{F15.6/ES15.8}       & \texttt{\%15.6f/\%15.8E}    & Frequency wavenumber in \cm\ or wavelength in nm or \um \\
$I$ or $\epsilon$                  & \texttt{ES15.8}             & \texttt{\%15.8E}            & Absorption or emission intensity in cm molecule$^{-1}$ or erg (s molecule sr)$^{-1}$ \\
$J'$                               & \texttt{I7/F7.1}            & \texttt{\%7d/\%7.1f}        & $J$-quantum number (integer/half-integer) for upper state \\
$E'$                               & \texttt{F12.6}              & \texttt{\%12.6f}            & Upper state energy in \cm \\
$J''$                              & \texttt{I7/F7.1}            & \texttt{\%7d/\%7.1f}        & $J$-quantum number (integer/half-integer) for lower state \\
$E''$                              & \texttt{F12.6}              & \texttt{\%12.6f}            & Lower state energy in \cm \\
(QN$'$)                            & See definition file & See definition file & Quantum numbers for upper state (optional) \\
(QN$''$)                           & See definition file & See definition file & Quantum numbers for lower state (optional) \\
\midrule
\multicolumn{4}{c}{\textbf{Cross sections \texttt{.xsec} and plots \texttt{.png}}} \\
\multicolumn{4}{c}{Fortran format: (\texttt{F15.6/ES15.8,1x,ES15.8})} \\
\midrule
$\tilde{\nu}$ or $\tilde{\lambda}$ & \texttt{F15.6/ES15.8}  & \texttt{\%15.6f/\%15.8E}  & Frequency wavenumber in \cm\ or wavelength in nm or \um \\
$\sigma$ & \texttt{ES15.8} & \texttt{\%15.8E} & Absorption or emission cross sections in cm$^2$ molecule$^{-1}$ or erg cm (s molecule sr)$^{-1}$ \\
\bottomrule
\end{tabular}
\begin{tablenotes}[flushleft]
\footnotesize
\item For oscillator strengths, stick spectra, and cross sections, plots can be generated with the x-axis in wavenumber (\cm) or wavelength (nm or \um), and with the y-axis on either a linear or logarithmic scale. 
\item Quantum number labels and formats are defined in \ExoMol\ definition \texttt{.def.json} file or \ExoAtom\ definition \texttt{.adef.json} file. 
\item $J$: Total angular momentum quantum, excluding nuclear spin. 
\item *: Lifetimes are stored in a column placed immediately after the uncertainty column. Where a lifetime column is already present, it is overwritten; otherwise, a new lifetime column is added. All remaining columns are kept formats unchanged from the original \ExoMol\ or \ExoAtom\ states file. 
\item $\tilde{\nu}$: It is wavenumber in unit \cm\ and saved in format \texttt{F15.6} and \texttt{\%15.6f}. 
\item $\tilde{\lambda}$: It is wavelength in unit nm or \um\ and saved in format \texttt{ES15.8} and \texttt{\%15.8E}. 
\end{tablenotes}
\end{threeparttable}
\end{table}
\begin{table}
\centering
\caption{Common spectroscopic quantities and symbols used throughout \PyExoCross.}
\label{tab:notation}
\setlength{\tabcolsep}{8mm}
\begin{tabular}{lll}
\toprule
Symbol & Description & Units \\
\midrule
$'$ & Upper state & \\
$''$ & Lower state & \\
$T$ & Temperature & K \\
$T_{\textrm{vib}}$ & Vibrational temperature & K \\
$T_{\textrm{rot}}$ & Rotational temperature & K \\
$P$ & Pressure & bar \\
$\tilde{E}_n$ & Energy term value, $\tilde{E}_n=E_n/hc$ & \cm \\
$\Delta E$ & Energy uncertainty & \cm \\
$\Delta \nu$ & Transition uncertainty & \cm \\
$\g_n^{\rm tot}$ & Total degeneracy, $\g_n^{\mbox{\scriptsize{tot}}}=\g_n^{\mbox{\scriptsize{ns}}}(2J_n+1)$ & \\
$\g_n^{\mbox{\scriptsize{ns}}}$ & Nuclear-spin statistical weight factor &  \\ 
$J$ and $J_n$ & Total angular momentum quantum number &  \\
$v$ & Generic vibrational (vibronic) quantum numnber & \\
$k$ & Rotational quantum number & \\
$h$ & Planck constant & erg$\cdot$s \\
$c$ & Speed of light & cm$/$s \\
$k_B$ & Boltzmann constant & erg$/$K \\
$c_2$ & Second radiation constant, $c_2=hc/k_B$ & cm$\cdot$K \\
$R$ & Molar gas constant & J K$^{-1}$ mol$^{-1}$ \\
$A$ & Einstein $A$-coefficient & s$^{-1}$ \\
$\tilde{\nu}$ & Transition wavenumber $\tilde{\nu}=|\tilde{E}'-\tilde{E}''|$ & \cm \\
$\Delta \tilde{\nu}$ & Signed displacement from the line centre $\Delta\tilde{\nu} = \tilde{\nu}-\tilde{\nu}_{\textrm{centre}}$ & \cm \\ 
$Q(T)$ & Partition function & \\
$C_p(T)$ & Partition function & J K$^{-1}$ mol$^{-1}$ \\
$W(T)$ & Cooling function & erg (s molecule sr)$^{-1}$ \\
$\tau$ & Radiative lifetime & s \\
$\gf$ & Weighted oscillator strength & \\
$f$ & Actual oscillator strength & \\
$fi$ or $f \leftarrow i$ & Absorption transition from the initial state $i$ to the final state $f$ & \\
$if$ or $i \leftarrow f$ & Emission transition from the final state $f$ to the initial state $i$ & \\
$I$ and $I_{fi}$ & Absorption coefficient or intensity & cm molecule$^{-1}$ \\
$\epsilon$ and $\epsilon_{if}$ & Emission coefficient or emissivity & erg (s molecule sr)$^{-1}$ \\
$S$ & Intensity or emission & cm molecule$^{-1}$ or erg (s molecule sr)$^{-1}$ \\
$f_{\tilde{\nu}}(\tilde{\nu})$ & Line profile & \\
$\sigma_{\textrm{ab}}$ and $\sigma_{fi}$ & Absorption cross sections & cm$^2$ molecule$^{-1}$ \\
$\sigma_{\textrm{em}}$ and $\sigma_{if}$ & Emission cross sections & erg cm (s molecule sr)$^{-1}$ \\
$F(T)$ & Population & \\
$n_{\textrm{vib}}$ & Non-LTE vibronic population fraction (vibrational density) & \\
vib & Vibrational (vibronic)& \\
rot & Rotational& \\
\bottomrule
\end{tabular}
\end{table}

\subsubsection{Data format conversion} \label{sec:convert}

A key practical requirement in spectroscopic post-processing is the ability to move between database formats without manually rewriting line list structures. \PyExoCross\ therefore includes data conversion functionality that allows users to import and export line list information between the principal supported data formats. 

The conversion routines are designed to preserve the quantities that are required for subsequent spectroscopic calculations, including transitions radiative parameters, states information, and the quantum labels needed for filtering or interpretation. Since different databases do not always provide identical metadata or adopt the same field definitions, the conversion step is not treated as a purely mechanical reformatting operation. Instead, it involves explicit metadata interpretation and remapping, especially for species labels and quantum number descriptors.
This conversion functionality provides a more flexible computational pipeline for spectroscopic analysis, especially in projects that require cross-comparison of multiple databases.

As shown in Table~\ref{tab:conv}, the currently implemented conversion pathways are centred on interoperability between the ExoMol-related formats and HITRAN format representations. In particular, \PyExoCross\ supports conversion from \ExoMol, \ExoMolHR, and \ExoAtom\ to HITRAN format, as well as reverse conversion from HITRAN/HITEMP to \ExoMol\ format.
\begin{table}
\centering
\caption{Supported data format conversion relationships between the spectroscopic databases currently interfaced in \PyExoCross.}
\label{tab:conv}
\setlength{\tabcolsep}{7.5mm}
\begin{tabular}{ccccc}
\toprule
Source database & Source data format & Source file format & Target data format & Target file format \\
\midrule
\multicolumn{5}{c}{\textbf{Molecular database}} \\
HITRAN/HITEMP & HITRAN    & \texttt{.par}        & \ExoMol & \texttt{.states.bz2} \\
              &           &                      &         & \texttt{.trans.bz2}  \\
              &           &                      &         & \texttt{.broad}      \\
\ExoMol       & \ExoMol   & \texttt{.states.bz2} & HITRAN  & \texttt{.par}        \\
              &           & \texttt{.trans.bz2}  &         &                      \\
              &           & \texttt{.broad}      &         &                      \\
\ExoMolHR     & \ExoMolHR & \texttt{.csv/.parquet}        & HITRAN  & \texttt{.par}        \\
\midrule
\multicolumn{5}{c}{\textbf{Atomic database}} \\
\ExoAtom      & \ExoMol   & \texttt{.states}     & HITRAN  & \texttt{.par}        \\ 
              &           & \texttt{.trans}      &         &                      \\
              &           & \texttt{.broad}      &         &                      \\
\bottomrule
\end{tabular}
\end{table}

\subsubsection{Thermodynamic and radiative quantities} \label{sec:thermo_rad}

In addition to spectral synthesis, \PyExoCross\ provides a set of derived
thermodynamic and radiative quantities, including partition functions (see Eq.~(\ref{eq:pf}) below),
specific heats (see Eq.~(\ref{eq:cp})), cooling functions (see Eq.~(\ref{eq:cf})), radiative lifetimes (see Eq.~(\ref{eq:life})), and oscillator
strengths (see Eq.~(\ref{eq:os})); see discussion given by \citet{jt914}. These quantities are evaluated within the same unified framework used for the other post-processing tasks, so that species definitions, state
energies, degeneracies, transition probabilities, and associated metadata are handled consistently across workflows. 
This design allows the same underlying spectroscopic data to support not only intensity and cross-section calculations, but also the derivation of auxiliary quantities that are frequently required in thermodynamic analysis, radiative modelling, and database validation.

Operationally, these quantities are obtained either directly from the internal state- and transition-based representation or, where available, from database-provided tabulations. 
Their implementation within the common software architecture makes them straightforward to compute, inspect, and reuse in scripted or batch workflows, particularly over large temperature grids. Variables used in the following equations are defined in Table~\ref{tab:notation}.

\paragraph{Partition functions $\textsl{Q}(T)$}
Partition functions quantify the thermal population of spectroscopic states and provide the normalization required in intensity calculations.
\begin{equation} 
    \textsl{Q}(T)=\sum_n \g_n^{\mbox{\scriptsize{tot}}}e^{-c_2\tilde{E}_n/T}.
    \label{eq:pf}
\end{equation}

\paragraph{Specific heats $\textsl{C}_p(T)$}
Specific heats are obtained from the partition function $Q(T)$ and its first two temperature derivatives.
\begin{equation}
    \textsl{C}_p(T) = R\left [\frac{\textsl{Q}''}{\textsl{Q}}-\left (\frac{\textsl{Q}'}{\textsl{Q}} \right )^2 \right ]+\frac{5R}{2}.
    \label{eq:cp}
\end{equation}

\paragraph{Cooling functions $\textsl{W}(T)$}
Cooling functions are evaluated from the radiative contributions of individual transitions.
\begin{equation} 
    \textsl{W}(T) = \frac{1}{4 \pi \textsl{Q}(T)} \sum_{f,i} A_{fi} h c \tilde{\nu}_{fi} \g'_f e^{-c_2 \tilde{E}'_f / T}.
    \label{eq:cf}
\end{equation}

\paragraph{Radiative lifetimes $\tau$}
Radiative lifetimes measure the total spontaneous decay rates of each upper state.
\begin{equation} 
    \tau_i = \frac{1}{{\textstyle \sum_{f} A_{fi}}}.
    \label{eq:life}
\end{equation} 

\paragraph{Oscillator strengths $\gf$ or $f$}
Oscillator strengths are calculated from transition frequencies and Einstein $A$-coefficients. The weighted oscillator strength is $\gf$, and the actual oscillator strength is denoted as $f$.
\begin{equation}
\label{eq:os}
    \gf=\frac{\g'_\textrm{tot}A_{fi}}{(c\tilde{v}_{fi})^2} \quad \textrm{or} \quad f= \gf/\g''_{\textrm{tot}}.
\end{equation}

\subsubsection{Spectral synthesis} \label{sec:synthesis}

A central functionality of \PyExoCross\ is the synthesis of spectra from state-resolved and transition-resolved spectroscopic data. Within the common internal framework, transition frequencies and intensities are first assembled into stick spectra (see Section \ref{sec:lte_nlte}), which provide the discrete, unbroadened representation of the spectrum \citep{jt914}. 
These stick spectra may then be converted into broadened cross sections (see Eqs.~(\ref{eq:xsec})) by applying a selected line-shape model over a user-defined spectral grid \citep{jt914}. 
In the updated implementation, both stick spectrum and cross sections can be generated in wavenumber (\cm) or wavelength (nm or \um). The selected coordinate system is used consistently for the input range, output files, and cross-section grid. 
This workflow is used consistently across the supported databases and calculation modes, and forms the basis for both LTE and non-LTE spectral simulations.

\paragraph{Stick spectra $I$ or $\epsilon$}
Stick spectra constitute the discrete spectral representation prior to any profile broadening and are available in \PyExoCross\ for both absorption and emission calculations. They may be generated under either LTE or non-LTE conditions, with the corresponding transition intensities evaluated according to the relevant formulations described in Section~\ref{sec:lte_nlte}.

\paragraph{Cross sections $\sigma$}
Broadened cross sections are obtained by distributing the integrated intensity of each discrete transition over frequency space using a normalized line-shape function centred at the transition position. In \PyExoCross, this procedure is applied to both absorption and emission spectra for both LTE and non-LTE conditions. 
The absorption cross sections $\sigma_{\textrm{ab},fi}$  and emission cross sections $\sigma_{\textrm{em},if}$ at an absolute wavenumber $\tilde{\nu}$ are determined bu their integrated intensities and line profiles. For a transition with positive line centre wavenumber $\tilde{\nu}_{fi}=E_f-E_i$, and transitions with negative values are discarded. We define the signed displacement from the line centre as $\Delta\tilde{\nu} = \tilde{\nu}-\tilde{\nu}_{fi}$.
The cross sections are then expressed as 
\begin{equation}
\label{eq:xsec}
\begin{aligned}
    I_{fi}
    &=
    \int_{-\infty}^{\infty}
    \sigma_{\mathrm{ab},fi}
    \left(\tilde{\nu}\right)
    \mathrm{d}\Delta\tilde{\nu} \quad \quad \Rightarrow
    &
    \sigma_{\mathrm{ab},fi}(\tilde{\nu})
    &=
    I_{fi}f_{fi}(\Delta\tilde{\nu}),
    \\
    \epsilon_{if}
    &=
    \int_{-\infty}^{\infty}
    \sigma_{\mathrm{em},if}
    \left(\tilde{\nu}\right)
    \mathrm{d}\Delta\tilde{\nu}  \quad \quad \Rightarrow
    &
    \sigma_{\mathrm{em},if}(\tilde{\nu})
    &=
    \epsilon_{if}f_{if}(\Delta\tilde{\nu}),
\end{aligned}
\end{equation}
where the line profiles are normalized according to
\begin{equation}
    \int_{-\infty}^{\infty}
    f_{fi}(\Delta\tilde{\nu})\,
    \mathrm{d}\Delta\tilde{\nu}
    =
    \int_{-\infty}^{\infty}
    f_{if}(\Delta\tilde{\nu})\,
    \mathrm{d}\Delta\tilde{\nu}
    =
    1.
\end{equation}
Although the normalized line profiles are formally defined over an infinite spectral interval, their numerical evaluation is restricted to a finite region around each transition centre, see \citet{jt909}. The treatment of line wing cut-off follows the implementation described for the original version of \PyExoCross\ \citep{jt914}. 
For a transition centred at $\tilde{\nu}_{fi}$, the profile contribution is evaluated only at grid points satisfying
$
\left|\tilde{\nu}-\tilde{\nu}_{fi}\right|
\leq \Delta\tilde{\nu}_{\mathrm{cut}},
$
where $\Delta\tilde{\nu}_{\mathrm{cut}}$ is the user-defined line wing cut-off. Transitions lying outside the requested output interval may therefore still be included when their truncated wings overlap the spectral grid. The same truncation criterion is applied in the command line and Python package workflows and across the supported computational backends.

\begin{figure}
\centering
\begin{subfigure}{0.49\textwidth}
    \centering
    \includegraphics[width=\textwidth]{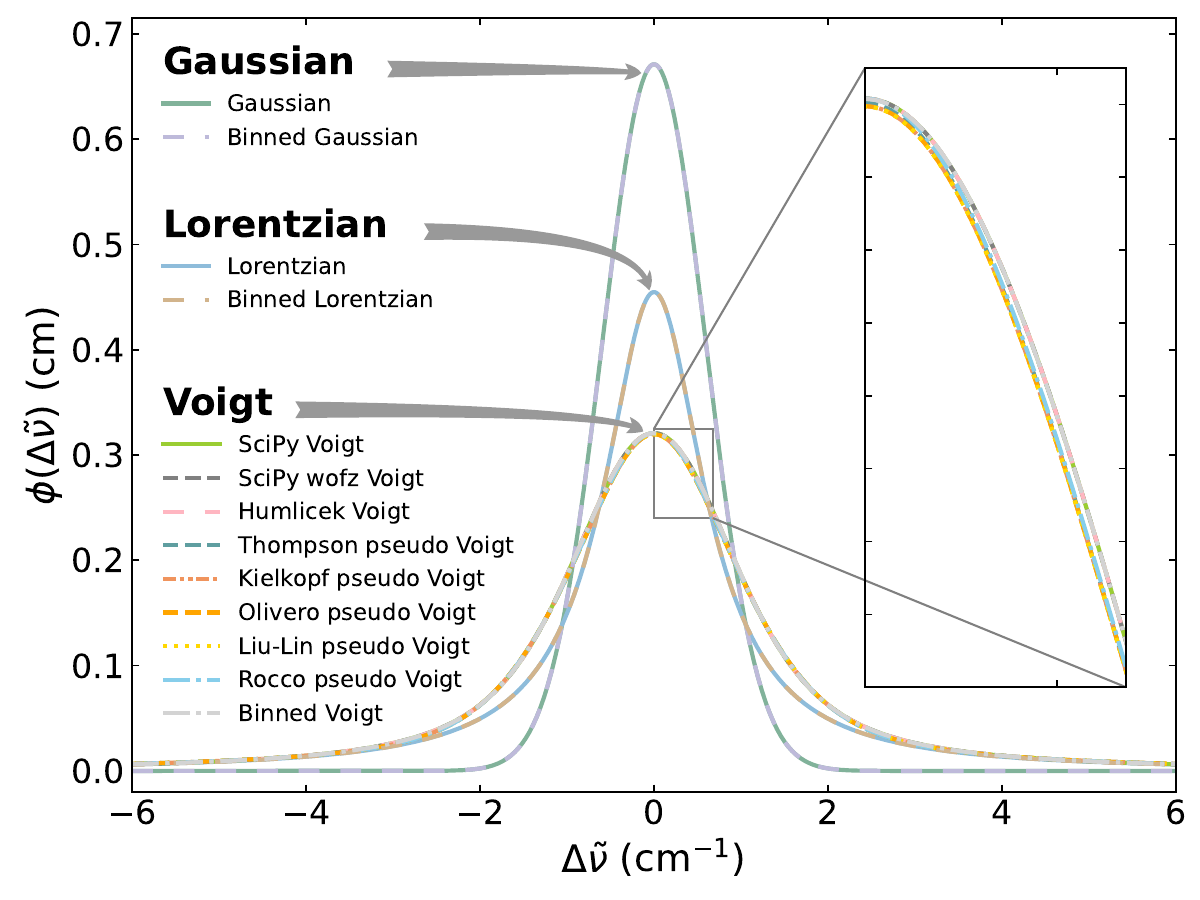}
    \caption{Compare all line profiles supported in the \PyExoCross}
\end{subfigure}
\begin{subfigure}{0.49\textwidth}
    \centering
    \includegraphics[width=\textwidth]{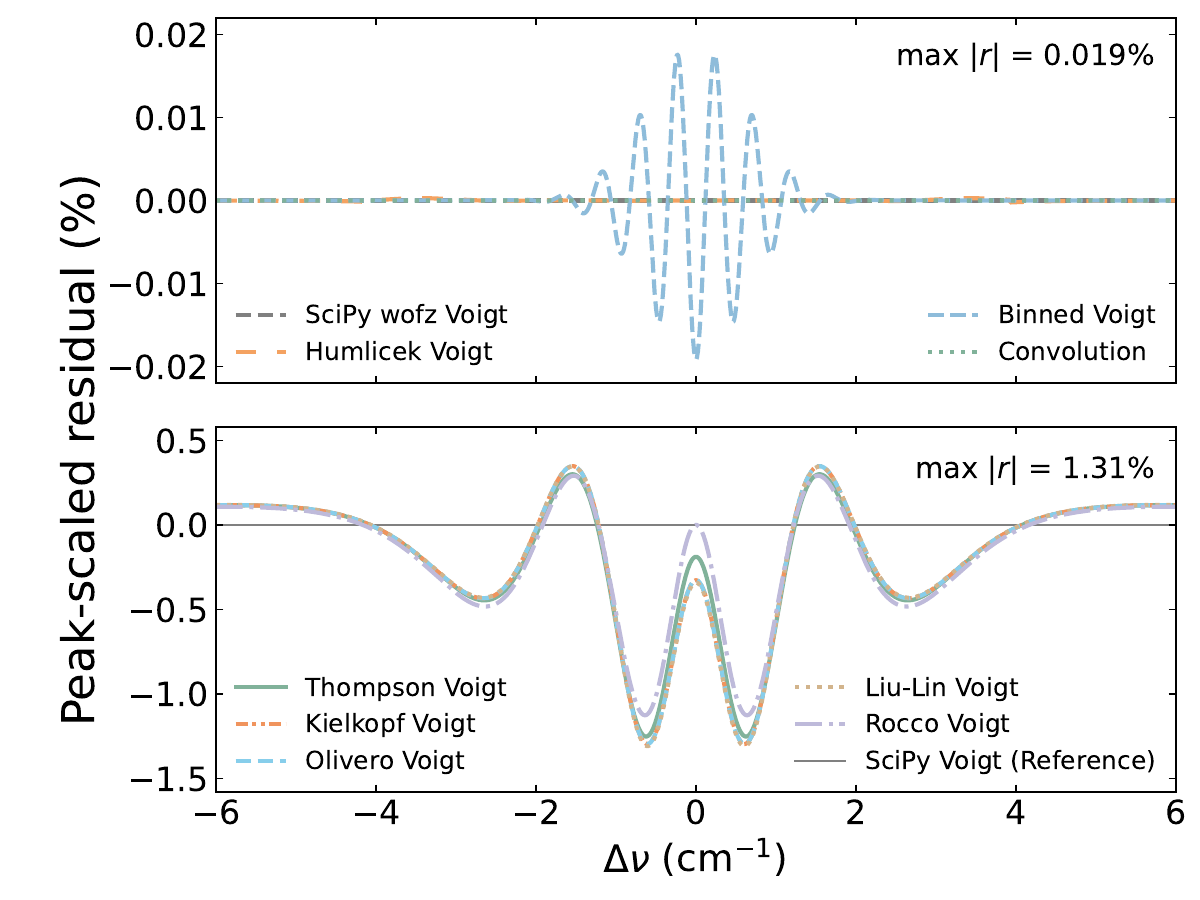}
    \caption{Peak-scaled residuals relative to the SciPy Voigt profile}
\end{subfigure}
\caption{Comparison of line profiles implemented in \PyExoCross. All profiles are normalized to unit area with the same integrated line intensity.}
\label{fig:GLV}
\end{figure}

\PyExoCross\ supports the standard profile models commonly used in spectroscopic calculations, including sampling and binned methods of Doppler (Gaussian), Lorentzian, and Voigt line profiles. 
Details for a total of 15 line profile methods are introduced in \citet{jt914}, see Table~\ref{tab:profile}. 
Fig.~\ref{fig:GLV} compares all profiles supported in the \PyExoCross. These broadened representations provide the final cross sections on the user-defined spectral grid.

\PyExoCross\ can also incorporate lifetime broadening into Voigt line profiles to calculate predissociation-broadened cross sections using the upper state lifetimes provided in the \ExoMol\ states files \citep{jt914}. This treatment applies to discrete transitions involving predissociative bound or quasi-bound states \citep{jt898}.

\section{Simulating LTE and non-LTE spectra} \label{sec:nlte}

\begin{table}
\centering
\caption{Methods used for evaluating Gaussian, Lorentzian, and Voigt-type line profiles in \PyExoCross.}
\label{tab:profile}
\setlength{\tabcolsep}{5.2mm}
\begin{tabular}{lll}
\toprule
Method & Description & Reference \\
\midrule
\multicolumn{3}{c}{\textbf{Direct profiles}} \\
Doppler & Determined by temperature, line position, and molecular mass & \citep{squires2001practical} \\
Binned Doppler & Bin-integrated Doppler profile & \citep{jt542} \\
Gaussian & User-defined half-width at half-maximum & \citep{squires2001practical} \\
Binned Gaussian & Bin-integrated Gaussian profile & \citep{jt542} \\
Lorentzian & For pressure broadening & \citep{feller1971introduction,johnson1994continuous} \\
Binned Lorentzian & Bin-integrated Lorentzian profile & \citep{jt708} \\
\midrule
\multicolumn{3}{c}{\textbf{Voigt evaluations}} \\
SciPy Voigt & \texttt{scipy.special.voigt\_profile} & \citep{SciPy} \\
SciPy \texttt{wofz} Voigt & Evaluated from the real part of the Faddeeva function via & \citep{Faddeeva1961TablesOV,pierluissi1977fast} \\
& \texttt{scipy.special.wofz} & \citep{SciPy} \\
Huml\'i\v{c}ek Voigt & Huml\'i\v{c}ek rational approximation & \citep{humlivcek1979efficient,humlivcek1982optimized,kuntz1997new} \\
Binned Voigt & Bin-integrated Voigt profile & \citep{abramowitz1972handbook} \\
& & \citep{jt708,jt914} \\
\midrule
\multicolumn{3}{c}{\textbf{Pseudo-Voigt approximations}} \\
Thompson pseudo-Voigt & Thompson parametrization & \citep{thompson1987rietveld} \\
Kielkopf pseudo-Voigt & Kielkopf prescription & \citep{kielkopf1973new} \\
Olivero pseudo-Voigt & Olivero-Longbothum width formula & \citep{olivero1977empirical} \\
Liu-Lin pseudo-Voigt & Liu-Lin formulation & \citep{liu2001simple} \\
Rocco pseudo-Voigt & Di Rocco-Cruzado formulation & \citep{di2012voigt} \\
\bottomrule
\end{tabular}
\end{table}

\PyExoCross\ is designed to compute atomic and molecular absorption and emission spectra, including both stick spectra and cross sections, over customised grids of temperature and pressure using a variety of line profile models \citep{jt914} under both local thermodynamic equilibrium (LTE) and non-local thermodynamic equilibrium (non-LTE) conditions. In this section, we describe the new functionality implemented in \PyExoCross\ to enable the calculation of non-LTE absorption and emission line intensities and cross sections.

In previous studies of exoplanet atmospheres, atmospheric species are generally assumed to satisfy local thermodynamic equilibrium. 
Under this assumption, state populations are determined by the local kinetic temperature and follow a Boltzmann distribution. However, non-LTE effects are known to occur in a range of planetary and astrophysical environments and can lead to substantial departures from LTE spectral features \citep{22WrWaYu.nLTE}. Observational evidence for non-LTE processes has been reported for the atmospheres of the Earth and other Solar System planets \citep{01LoTaxx}, including Venus \citep{LopezValverde2007}, Mars \citep{LopezValverde2005}, and the gas giant planets \citep{Kim2015}. 
Similar effects are also encountered in stellar atmospheres, comets \citep{Weaver1984}, and the interstellar medium \citep{vanderTak2007,Lique2008}. 
Recent non-LTE radiative transfer modelling of exoplanet atmospheres has shown that non-LTE molecular populations can substantially alter opacities, observable spectra, photodissociation rates, and retrieved molecular abundances \citep{jt995}.
The theoretical foundation of non-LTE spectroscopy dates back to the work of \citet{milne1930handbuch} and has subsequently been developed through a series of significant studies \citep{Curtis1956,Houghton1969,Kuhn1969,Dickinson1972,Kumer1974,LpezPuertasI1986,LpezPuertasII1986,Wintersteiner1992}.

Maser (Microwave Amplification by Stimulated Emission of Radiation) emission is a well-documented astrophysical phenomenon that occurs under non-LTE conditions. 
Unlike thermal radiation, which follows the Boltzmann distribution of level populations under LTE, maser emission results from population inversion and the consequent stimulated amplification of radiation at specific transitions \citep{Elitzur1992,Gray2012}. 
Maser processes are commonly observed in astrophysical environments such as star-forming regions, circumstellar envelopes of evolved stars, and galactic nuclear regions \citep{Reid1981,Humphreys2007}. 
The comprehension of maser emission in non-LTE conditions continues to be a dynamic field of inquiry, with persistent endeavours to enhance models and integrate supplementary physical mechanisms.

Atoms and molecules may be driven to non-LTE states by specific physical features of some exoplanetary atmospheres. 
Such as, non-LTE state distributions emerge from atoms and molecules losing energy mainly in their rotational mode rather than their vibrational mode in high altitude shock regions at the edges of fast moving jet streams. 
Stellar pumping selectively energises vibrational modes of atoms and molecules in areas of the atmosphere exposed to intense stellar radiation, forcing them into non-LTE states.
The non-LTE atomic and molecular states produce emission and transmission spectra that significantly differ from their LTE equivalents, and these discrepancies can be observed under appropriate conditions.

\subsection{Populations and densities} \label{sec:pop_den}

\PyExoCross\ can simulate non-LTE spectra using a simple two-temperature approach which is between rotational and vibrational (vibronic) temperatures when computing intensities, partition functions, which is generally known as the Treanor approximation \citep{68TrRiRe.nLTE}, or other temperature-dependent features \citep{jt708}. 
Different electrical, vibronic, or vibrational bands can be individually simulated using an effective filtering system based on quantum numbers.
For this purpose, the total energy is estimated as the sum of the vibrational (or vibronic) and rotational components \citep{68TrRiRe.nLTE}.
\begin{equation}
    \tilde{E}_{v,J,k}=\tilde{E}_v^{\textrm{vib}}+\tilde{E}_{J,k}^{v,\textrm{rot}},
    \label{eq:Enonlte}  
\end{equation}
where $v$ is generic vibrational (vibronic) quantum numbers, $J$ is total angular momentum, and $k$ is rotational quantum numbers. 

The pure vibronic contributions are considered as the corresponding energy values at $J=0$ for integer spins, $J=1/2$ for non-integer (half-integer) spins, or the lowest $J$ permitted by the symmetry of the electronic term and parity, typically associated with the lowest states (usually `$+$' or `e') \citep{jt708}.  
In some circumstances, certain spectral lines include NaN values for specified quantum numbers in the \texttt{.states} file. The program \PyExoCross\ normally ignores these lines during processing.
The rotational contribution is simply given by: 
\begin{equation}
    \tilde{E}_{J,k}^{v,\textrm{rot}}=\tilde{E}_{v,J,k}-\tilde{E}_v^{\textrm{vib}}.
    \label{eq:Erot nonlte}  
\end{equation}

The spectral simulations are conducted under three different situations, as summarized in Table~\ref{tab:3 cases LTE NLTE}.
\begin{table}
\centering
\caption{Summary of the three cases considered for spectral simulations.}
\label{tab:3 cases LTE NLTE}
\setlength{\tabcolsep}{5.2mm}
\begin{tabular}{llccccll}
\toprule
Case & Distribution & $T_{\textrm{rot}}$ & $T_{\textrm{vib}}$ & Relation & Section & Reference \\
\midrule
LTE     & Boltzmann distribution & \cmark & \cmark & $T=T_{\textrm{rot}}=T_{\textrm{vib}}$   & Section \ref{sec:lte pop}          & \citep{landau2013statistical} \\
Non-LTE & Treanor distribution   & \cmark & \cmark & $T_{\textrm{rot}}\neq T_{\textrm{vib}}$ & Section \ref{sec:nlte pop 2T}      & \citep{68TrRiRe.nLTE} \\
Non-LTE & Vibronic distribution  & \cmark & \xmark &                                         & Section \ref{sec:nlte pop density} & \citep{74Berryx} \\
\bottomrule
\end{tabular}
\end{table}

\subsubsection{LTE populations using the Boltzmann distributions} \label{sec:lte pop}

In the LTE case, the rotational and vibrational temperatures are equivalent, and the populations of molecular states follow the Boltzmann distributions at a specified temperature $T$, as indicated by \citep{22WrWaYu.nLTE}:
\begin{equation}
    F_{J,v,k}(T)=\frac{\g_{J,v,k}^{\textrm{tot}}e^{-c_2\tilde{E}_{J,v,k}/T}}{Q(T)},
    \label{eq:F lte}  
\end{equation}
where $\g_{J,v,k}^{\textrm{tot}}$ is the total degeneracy and $\tilde{E}_{J,v,k}=E_{J,v,k}/hc$ is the state energy term value (\cm).
$c_2=hc/k_B$ is the second radiation constant (cm K) and $k_B$ is the Boltzmann constant (erg$/$K). T is the temperature in K and $Q(T)$ is the temperature-dependent partition function.
The LTE partition function $Q(T)$ is defined as a sum over all states, see Eq.~(\ref{eq:pf}).

\subsubsection{Non-LTE populations using two-temperature Treanor distributions} \label{sec:nlte pop 2T}

To broaden this framework to include both vibrational and rotational temperatures, we apply a simplified bi-temperature Treanor approximation model \citep{68TrRiRe.nLTE} and investigate the spectral characteristics of different vibrational bands that contribute to a generic non-LTE spectrum. In the Treanor approximation, vibrational populations are represented by a vibrational Boltzmann distribution at a distinct temperature $T_{\textrm{vib}}$ and rotational populations by $T_{\textrm{rot}}$. The total non-LTE population of a specific state is expressed as the multiplication of two Boltzmann distributions:
\begin{equation}
    F_{J,v,k}(T_{\textrm{vib}},T_{\textrm{rot}})=\frac{\g_{J,v,k}^{\textrm{tot}} e^{-c_2\tilde{E}_v^{\textrm{vib}}/T_{\textrm{vib}}}e^{-c_2\tilde{E}_{J,k}^{v,\textrm{rot}}/T_{\textrm{rot}}}}{Q(T_{\textrm{vib}},T_{\textrm{rot}})}.
    \label{eq:F nlte 2T}  
\end{equation}
This non-LTE partition function $Q(T_{\textrm{vib}},T_{\textrm{rot}})$ depends on two different temperatures $T_{\textrm{vib}}$ and $T_{\textrm{rot}}$.
\begin{equation}
    Q(T_{\textrm{vib}},T_{\textrm{rot}})=\sum_{J,v,k}\g_{J,v,k}^{\textrm{tot}}e^{-c_2\tilde{E}_{v}^{\textrm{vib}}/T_{\textrm{vib}}}e^{-c_2\tilde{E}_{J,k}^{v,\textrm{rot}}/T_{\textrm{rot}}}.
    \label{eq:Q nlte 2T}  
\end{equation}

\subsubsection{Non-LTE populations using custom vibronic densities} \label{sec:nlte pop density}

Non-LTE conditions can also involve the application of a vibronic distribution and a rotational temperature.
The vibrational densities 
The user-supplied quantities $n_v^{\textrm{vib}}$ specify the relative
weights assigned to the vibrational or vibronic manifolds, which can be applied as directly incorporated into the calculations through the custom provided \texttt{.states} file. 
The same weight is assigned to all rotational states belonging to a given manifold. 
The sum of the non-LTE vibronic population fraction (vibrational densities) $n_v^{\textrm{vib}}$ is normalized to one.
\begin{equation}
    \sum_i n_i^{\textrm{vib}}=1.
    \label{eq:Nvib}
\end{equation}
The rotational states are assumed to be populated according to a Boltzmann distribution corresponding to a rotational temperature $T_{\textrm{rot}}$, which also reflects the kinetic temperature of the environment \citep{LpezPuertas2001}. The total population of $J$, $v$, and $k$ is then approximated as the product of these distributions \citep{22WrWaYu.nLTE}.
\begin{equation}
    F_{J,v,k}(T_{\textrm{rot}})=\frac{\g_{J,v,k}^{\textrm{tot}}n_v^{\textrm{vib}}e^{-c_2\tilde{E}_{J,k}^{v,\textrm{rot}}/T_{\textrm{rot}}}}{{Q}(T_{\textrm{rot}})},
    \label{eq:F nlte Nvib}  
\end{equation}
where ${Q}(T_{\textrm{rot}})$ is the normalization function obtained by summing over all $J$, $v$, and $k$ states included in the calculation.
\begin{equation}
    Q(T_{\textrm{rot}})=\sum_{J,v,k}\g_{J,v,k}^{\textrm{tot}}n_v^{\textrm{vib}}e^{-c_2\tilde{E}_{J,k}^{v,\textrm{rot}}/T_{\textrm{rot}}}.
    \label{eq:Q nlte Nvib}  
\end{equation}

\subsection{LTE and non-LTE absorption and emission intensities} \label{sec:lte_nlte}

This section provides a more extensive presentation of the equations used to calculate absorption and emission intensities in both LTE and non-LTE conditions. 
It elaborates on the specific mathematical formulations for each case, highlighting the variations in the treatment of atomic and molecular energy state populations, temperature distributions, and the factors impacting radiative transfer in these two regimes.

\subsubsection{LTE absorption intensities} \label{sec:ab lte}

To simulate LTE absorption spectra, the Boltzmann distributions are applied to the expression for absorption line intensities \citep{96JoLaIw.CH,jt708}.
\begin{equation} 
\label{eq:abs intensity lte}
    I_{f \gets i} = \frac{\g'_{J,v,k} {A}_{fi}}{8 \pi c \tilde{\nu}^2_{fi}} \frac{e^{-c_2 \tilde{E}''_{J,v,k} / T} (1 - e^{-c_2 \tilde{\nu}_{fi} / T })}{Q(T)} 
    =\frac{\g'_{J,v,k}{A}_{fi}}{8 \pi c \tilde{\nu}^2_{fi}}\frac{1}{\g''_{J,v,k}}F_{J'',v'',k''}(T)(1-e^{-c_2\tilde{\nu}_{fi}/T}).
\end{equation}
$\g'_{J,v,k}$ and $\g''_{J,v,k}$ represent the total degeneracies of the upper and lower states, respectively. 
$F_{J'',v'',k''}(T)$ is expressed as  Eq.~(\ref{eq:F lte}) for lower state and can also be derived from the custom offered \texttt{.states} file.

\subsubsection{Non-LTE absorption intensities using two-temperature Treanor distribution} \label{sec:ab nlte 2T}

In the two-temperature Treanor distribution model, its partition function $Q(T_{\textrm{vib}},T_{\textrm{rot}})$ can be determined using  Eq.~(\ref{eq:Q nlte 2T}). 
\begin{equation}
\label{eq:nlte v}
    \tilde{\nu}_{fi} = \tilde{\nu}_{\textrm{vib}} + \tilde{\nu}_{\textrm{rot}} \quad \textrm{where} \quad 
    \tilde{\nu}_{\textrm{vib}} = \tilde{E}_{v}^{'\textrm{vib}} - \tilde{E}_{v}^{''\textrm{vib}} \quad \textrm{and} \quad 
    \tilde{\nu}_{\textrm{rot}} = \tilde{E}_{j,k}^{'v,\textrm{rot}} - \tilde{E}_{j,k}^{''v,\textrm{rot}}.
\end{equation}
\begin{equation}
\label{eq:nlte ab 2T}
    I_{f \gets i} = \frac{\g'_{J,v,k}{A}_{fi}}{8 \pi c \tilde{\nu}^2_{fi}} \frac{e^{-c_2 \tilde{E}_{v}^{''\textrm{vib}}/T_{\textrm{vib}}} e^{-c_2 \tilde{E}_{j,k}^{''v,\textrm{rot}}/T_{\textrm{rot}}} (1 - e^{-c_2 (\tilde{\nu}_{\textrm{vib}} / T_{\textrm{vib}}+\tilde{\nu}_{\textrm{rot}} / T_{\textrm{rot}}})}{Q(T_{\textrm{vib}},T_{\textrm{rot}})}.
\end{equation}
The upper state total degeneracy $\g'_{J,v,k}$ and the lower state energy $\tilde{E}''$ are used in this model.

\subsubsection{Non-LTE absorption intensities using custom vibrational (vibronic) density} \label{sec:ab nlte density}

A rotational temperature $T_{\textrm{rot}}$ is used in this method. The total degeneracy $\g'_{J,v,k}$ corresponds to the upper state, while the energy and the custom vibrational (vibronic) density $n''_{\textrm{vib}}$, as defined in  Eq.~(\ref{eq:Nvib}), corresponds to the lower states. The rotational temperature-dependent partition function $Q(T_{\textrm{rot}})$ is defined as Eq.~(\ref{eq:Q nlte Nvib}).
\begin{equation}
    \label{eq:nlte ab pop density}
    I_{f \gets i} = \frac{\g'_{J,v,k}{A}_{fi}}{8 \pi c \tilde{\nu}^2_{fi}} \frac{n''_{\textrm{vib}} e^{-c_2 \tilde{E}_{j,k}^{''v,\textrm{rot}}/T_{\textrm{rot}}} (1 - e^{-c_2 \tilde{\nu}_{fi} / T_{\textrm{rot}} })}{Q(T_{\textrm{rot}})}.
\end{equation}

\subsubsection{Non-LTE absorption intensities using custom rovibrational (rovibronic) population} \label{sec:ab nlte pop}

The non-LTE custom rovibrational (rovibronic) population $F_{J'',v'',k''}(T_{\textrm{vib}},T_{\textrm{rot}})$ in  Eq.~(\ref{eq:F nlte 2T}) for the lower state is applied in this case which is defined in  Eq.~(\ref{eq:F nlte 2T}). $\g'_{J,v,k}$ and $\g''_{J,v,k}$ are the total degeneracies for upper and lower states, respectively.
\begin{equation}
    \label{eq:nlte ab pop}
    I_{f \gets i} = \frac{\g'_{J,v,k}{A}_{fi}}{8 \pi c \tilde{\nu}^2_{fi}} \frac{1}{\g''_{J,v,k}}F_{J'',v'',k''}(T_{\textrm{vib}},T_{\textrm{rot}})(1 - e^{-c_2 \tilde{\nu}_{fi}/T}).
\end{equation}

\subsubsection{LTE emission intensities} \label{sec:em nlte}

To simulate LTE emission spectra, the Boltzmann distributions are applied to the expression for emission line intensities \citep{jt708}.
\begin{equation} 
\label{eq:ems intensity lte}
    \varepsilon_{i \gets f} = \frac{\g'_{J,v,k} hc {A}_{fi} \tilde{\nu}_{fi}} {4 \pi} \frac{e^{-c_2 \tilde{E}'_{J,v,k} / T}}{Q(T)} 
    = \frac{hc {A}_{fi} \tilde{\nu}_{fi}} {4 \pi} F_{J',v',k'}(T).
\end{equation}
$\g'_{J,v,k}$ and $\g''_{J,v,k}$ represent the total degeneracies of the upper and lower states, respectively. 
$F_{J',v',k'}(T)$ is expressed as  Eq.~(\ref{eq:F lte}) for upper state and can also be extracted from the custom offered \texttt{.states} file.

\subsubsection{Non-LTE emission intensities using two-temperature Treanor distribution} \label{sec:em nlte 2T}

In the two-temperature Treanor distribution model, its partition function $Q(T_{\textrm{vib}},T_{\textrm{rot}})$ can be determined using  Eq.~(\ref{eq:Q nlte 2T}). The upper state total degeneracy $\g'_{J,v,k}$ and energy $\tilde{E}'$ are used in the following equation.
\begin{equation}
    \label{eq:nlte em 2T}
    \varepsilon_{i \gets f} = \frac{hc}{4 \pi} \g'_{J,v,k}{A}_{fi}\tilde{\nu}_{fi} \frac{e^{-c_2 \tilde{E}_{v}^{'\textrm{vib}}/T_{\textrm{vib}}} e^{-c_2 \tilde{E}_{j,k}^{'\textrm{v,rot}}/T_{\textrm{rot}}}}{Q(T_{\textrm{vib}},T_{\textrm{rot}})}. 
\end{equation}

\subsubsection{Non-LTE emssion intensities using custom vibrational (vibronic) density} \label{sec:em nlte density}

In this method, total degeneracy $\g'_{J,v,k}$, energy $\tilde{E}'$, and custom vibrational (vibronic) density $n'_{\textrm{vib}}$ are all corresponded to the upper states, see  Eq.~(\ref{eq:Nvib}). The rotational temperature-dependent partition function $Q(T_{\textrm{rot}})$ is determined as  Eq.~(\ref{eq:Q nlte Nvib}).
\begin{equation}
    \label{eq:nlte em pop density}
    \varepsilon_{i \gets f} = \frac{hc}{4 \pi} \g'_{J,v,k}{A}_{fi} \tilde{\nu}_{fi} \frac{n'_{\textrm{vib}} e^{-c_2 \tilde{E}_{j,k}^{'\textrm{v,rot}}/T_{\textrm{rot}}}}{Q(T_{\textrm{rot}})}.
\end{equation}

\subsubsection{Non-LTE emission intensities using custom rovibrational (rovibronic) population} \label{sec:em nlte pop}

The upper state custom rovibrational (rovibronic) population $F_{J',v',k'}(T_{\textrm{vib}},T_{\textrm{rot}})$ in  Eq.~(\ref{eq:F nlte 2T}) can be derived from a user-provided \texttt{.states} file.
\begin{equation}
    \label{eq:nlte em pop}
    \varepsilon_{i \gets f} = \frac{hc}{4 \pi} {A}_{fi}\tilde{\nu}_{fi} F_{J',v',k'}(T_{\textrm{vib}},T_{\textrm{rot}}).
\end{equation}

\subsection{LTE and non-LTE comparison and discussion} \label{sec:lte_vs_nlte} 

We use NO and CO line lists from the \ExoMol\ database to illustrate two distinct effects of a two-temperature non-LTE distribution. The  temperatures adopted are illustrative and are chosen to separate rotational and vibrational population effects. They do not represent retrieved temperatures for any specific atmosphere. All cross
sections are calculated at $P=10^{-4}$ bar using the SciPy Voigt profile.

In the non-LTE calculations, $T_{\rm rot}$ is used as the kinetic-temperature proxy in the line profile calculation. It therefore controls the Doppler width and the temperature dependence of the pressure-broadening width. This treatment assumes that the translational and rotational temperatures are equal, while the vibrational population may be described by a separate temperature $T_{\rm vib}$.

For each molecule, the full range spectrum provides an overview of the line list coverage. The selected infrared region is then presented at a resolving power of $R=1000$, where $R=\lambda/\Delta\lambda \approx\tilde{\nu}/\Delta\tilde{\nu}$. The final panel shows the signed ratio $\sigma_{\rm NLTE}/\sigma_{\rm LTE}$. Ratios are displayed only where the LTE cross section exceeds $10^{-3}$ of its maximum value in the selected interval. This mask prevents ratios in negligible-opacity regions from dominating the comparison. The ratio represents a model spectral difference, rather than an uncertainty or error.

\subsubsection{Rotational population redistribution in NO}

\begin{figure*}
\centering
\begin{subfigure}{0.95\textwidth}
    \centering
    \includegraphics[width=\textwidth]{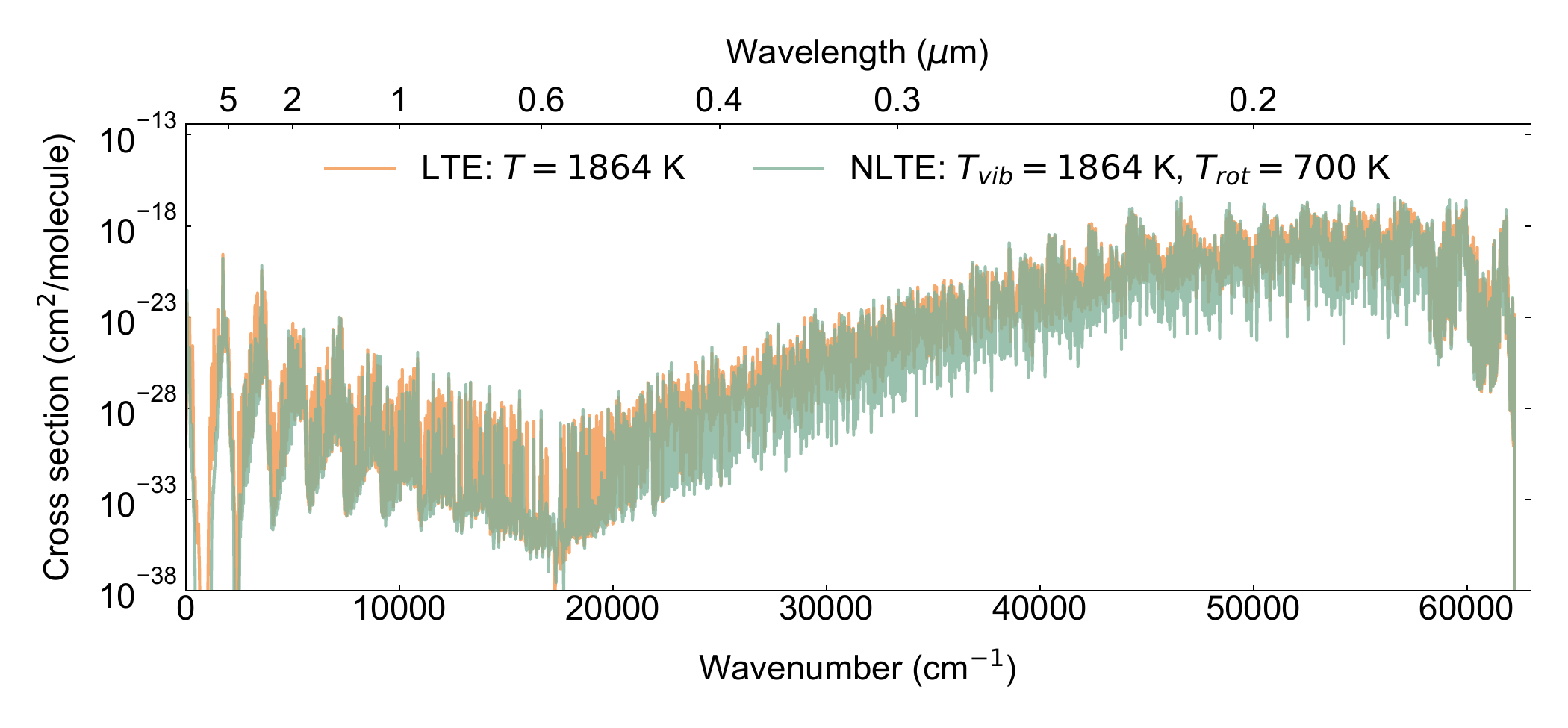}
\end{subfigure}
\begin{subfigure}{0.95\textwidth}
    \centering
    \includegraphics[width=\textwidth]{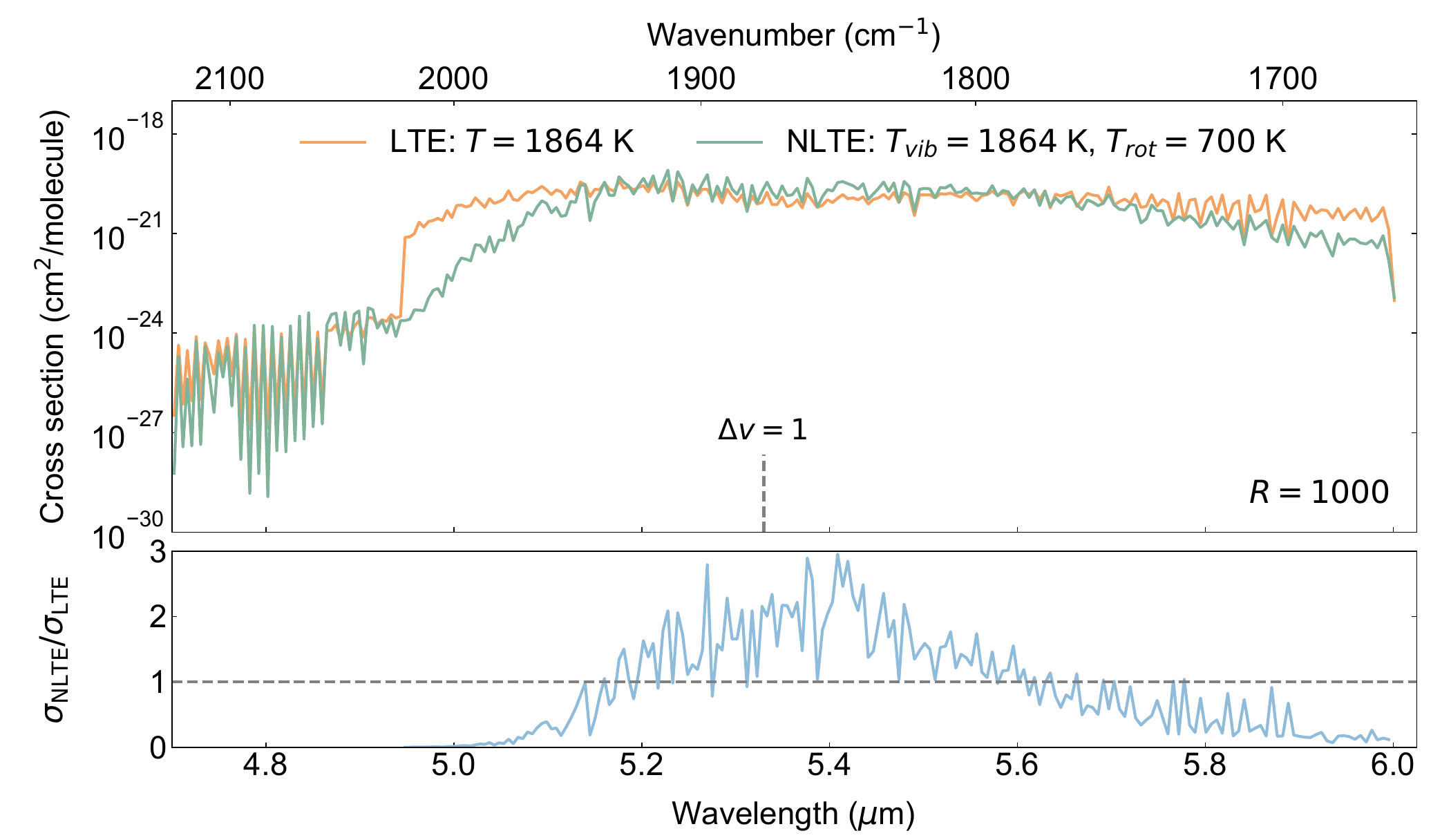}
\end{subfigure}
\caption{LTE and two-temperature non-LTE absorption cross sections of $^{14}$N$^{16}$O calculated from the XABC line list at
$P=10^{-4}$ bar. The LTE calculation uses $T=1864$ K, while the non-LTE calculation uses $T_{\rm vib}=1864$ K and $T_{\rm rot}=700$ K. The dotted marker indicates the approximate centre of the $\Delta v=1$ vibrational band sequence.}
\label{fig:lte_nlte_NO}
\end{figure*}

Figure~\ref{fig:lte_nlte_NO} compares LTE and non-LTE absorption cross sections calculated from the $^{14}$N$^{16}$O XABC line list \citep{jt831}. The LTE reference uses $T=1864$ K. In the non-LTE calculation, $T_{\rm vib}=1864$ K is retained while $T_{\rm rot}$ is reduced to $700$ K. This comparison therefore isolates the effect of rotational cooling without changing the vibrational temperature.

The full range spectra have the same transition coverage, but their opacity is redistributed because the lower rotational temperature favours lower-$J$ states and reduces the populations of highly excited rotational states. In the $4.7$--$6.0$ \um\ interval, this produces a contraction and reshaping of the rotational envelope around the $\Delta v=1$ vibrational band sequence. The non-LTE spectrum is consequently weaker in parts of the band wings, while smaller enhancements occur where opacity is transferred towards transitions involving lower rotational excitation. This directional change is shown explicitly by
$\sigma_{\rm NLTE}/\sigma_{\rm LTE}$ in Fig.~\ref{fig:lte_nlte_NO} bottom panel.

Because $T_{\rm vib}$ is unchanged in this example, the NO comparison is not intended to demonstrate enhanced vibrational hot bands. Its purpose is to show the rotational redistribution produced by $T_{\rm rot}<T_{\rm vib}$.

\subsubsection{Vibrational hot-band enhancement in CO}

\begin{figure*}
\centering
\begin{subfigure}{0.95\textwidth}
    \centering
    \includegraphics[width=\textwidth]
    {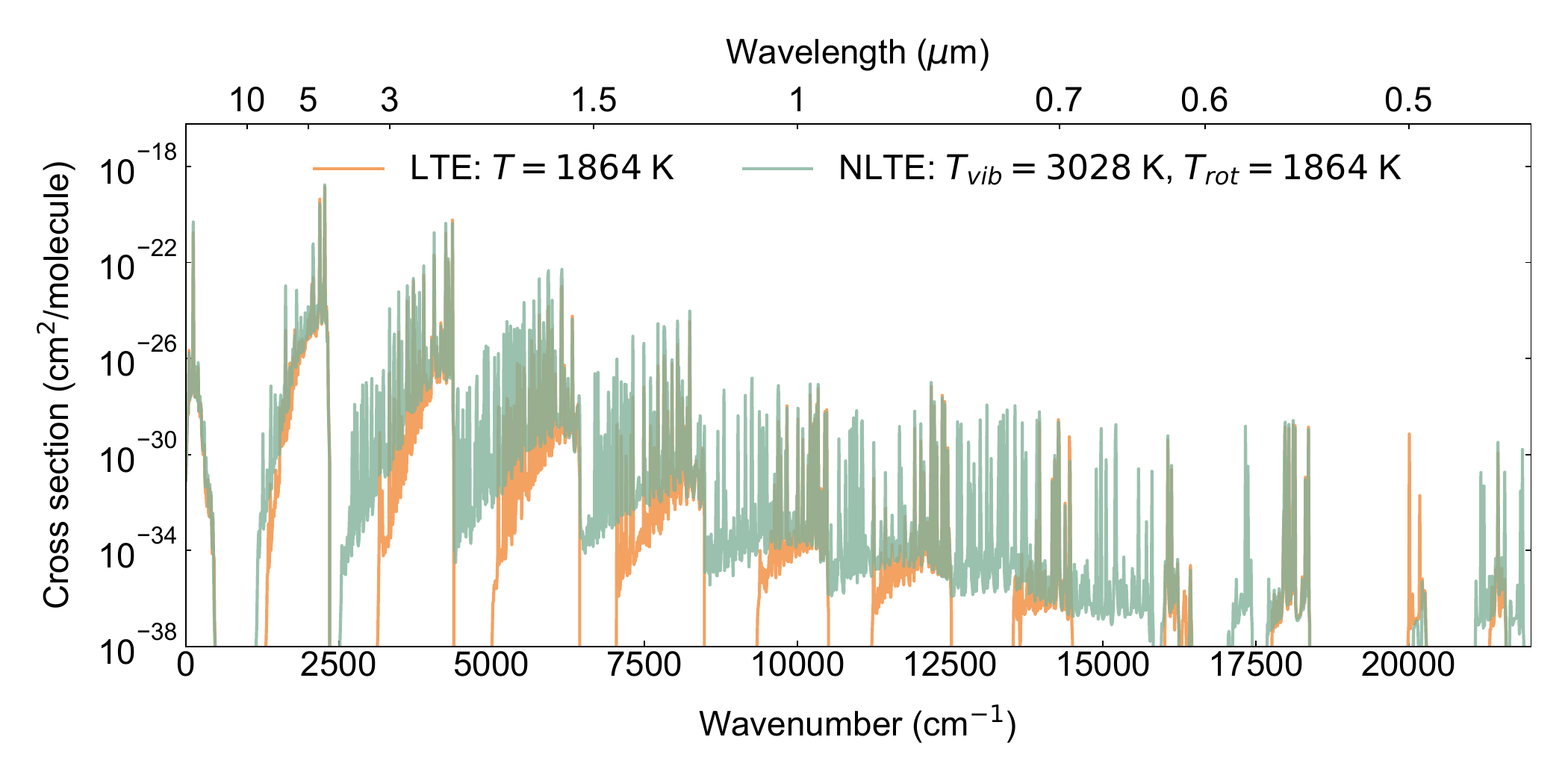}
\end{subfigure}
\begin{subfigure}{0.95\textwidth}
    \centering
    \includegraphics[width=\textwidth]
    {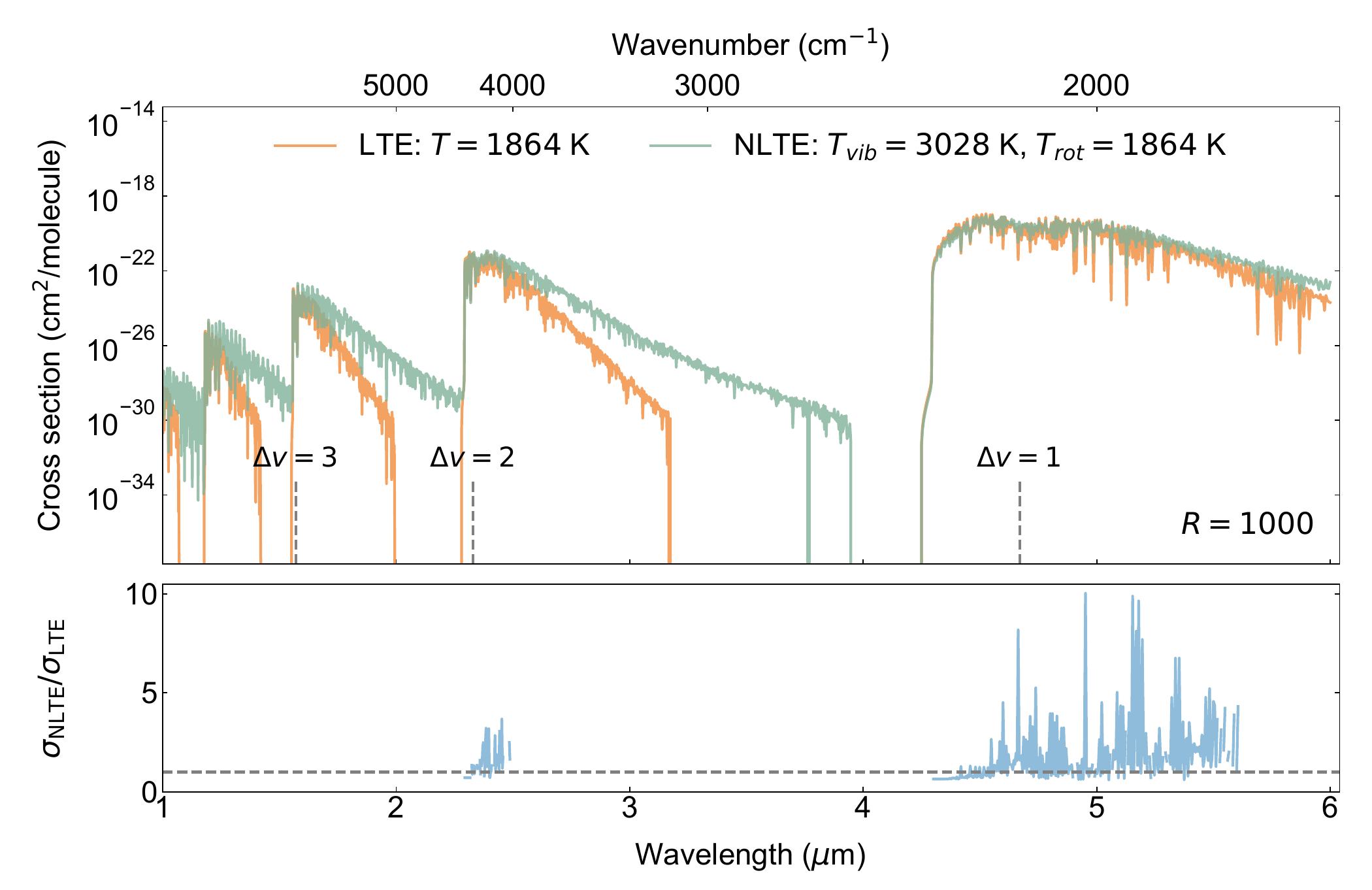}
\end{subfigure}
\caption{LTE and two-temperature non-LTE absorption cross sections of $^{12}$C$^{16}$O calculated from the Li2015 line list at $P=10^{-4}$ bar. The LTE calculation uses $T=1864$ K, while the non-LTE calculation uses $T_{\rm vib}=3028$ K and $T_{\rm rot}=1864$ K. The dotted markers indicate the approximate centres of the $\Delta v=1$, 2, and 3 vibrational band sequences.}
\label{fig:lte_nlte_CO}
\end{figure*}

Figure~\ref{fig:lte_nlte_CO} presents the complementary vibrational-heating case using the $^{12}$C$^{16}$O Li2015 line list \citep{15LiGoRo.CO}. The LTE reference again uses $T=1864$ K. For the non-LTE calculation, $T_{\rm rot}$ remains at $1864$ K while $T_{\rm vib}$ is increased to $3028$ K. The rotational distribution and line profile temperature are therefore unchanged, allowing the effect of the vibrational population to be identified more directly.

Increasing $T_{\rm vib}$ transfers population from the vibrational ground state to excited vibrational states. Transitions originating from these states form vibrational hot bands, which overlap the fundamental and overtone sequences and increase the opacity between and around their main band envelopes. The effect is particularly clear in the $\Delta v=2$ and $\Delta v=3$ regions, where the non-LTE spectrum retains substantially more opacity than the LTE spectrum away from the strongest band heads. The $\Delta v=1$ region is also reshaped, although increasing $T_{\rm vib}$ does not require every wavelength point to become stronger because the total population is redistributed among several vibrational states.

The dotted markers in Fig.~\ref{fig:lte_nlte_CO} bottom panel indicate the approximate centres of the $\Delta v=1$, 2, and 3 vibrational band sequences. They correspond approximately to the fundamental, first-overtone, and second-overtone regions, respectively. Each sequence contains many rovibrational transitions, including hot bands, so the markers do not represent individual spectral lines. The ratio in demonstrates both the magnitude and direction of the resulting changes.

The two examples demonstrate complementary non-LTE behaviour. Lowering $T_{\rm rot}$ at fixed $T_{\rm vib}$ primarily redistributes opacity within rotational band envelopes, as shown for NO. Increasing $T_{\rm vib}$ at fixed $T_{\rm rot}$ increases the populations of vibrationally excited states and makes hot-band structure more pronounced, as shown for CO.
The spectral response therefore depends on which population temperature departs from the LTE reference.

\section{Accuracy and performance} \label{sec:gpu}

\PyExoCross\ now provides three GPU execution modes to improve computational performance, especially for species with large datasets. GPU acceleration is currently implemented for the calculation of cooling functions, stick spectra, and cross sections.
The following tests evaluate the GPU implementation in terms of both numerical accuracy and computational performance, for individual spectroscopic components as well as for calculations carried out under LTE and non-LTE conditions. 
We also assess the performance of the combined stick spectra and cross sections workflow over multiple temperatures and pressures, as this represents a common use case in large-scale spectroscopic applications.

The performance evaluation also includes the data handling workflow used for large line lists. 
\PyExoCross\ reads transition files in chunks and uses a bounded streaming path for large \ExoMol\ files, avoiding the need to hold the full transition table in memory. 
For smaller inputs, transition chunks can be cached and reused across temperatures, pressures, and output products. 
In the combined stick spectra and cross sections workflow, each transition chunk is reused where possible for both outputs, reducing repeated file I/O, see Section~\ref{sec:stick and xsec}. 
Large output tables are also written in chunks. 
The reported timings therefore reflect both the accelerated numerical calculations and the practical cost of processing large spectroscopic datasets.

CPU reference calculations were performed on a Linux system with an AMD EPYC 7543 processor. CUDA benchmarks were conducted on the same system using a single NVIDIA GeForce RTX 4090 GPU with 24 GB memory. MPS (Metal Performance Shaders) accessed through the PyTorch MPS backend for GPU acceleration on Apple silicon, benchmarks were performed separately on macOS 15.1 using an Apple M1 Pro.

\subsection{GPU accuracy and performance for line profiles} \label{sec:gpu profile}

This section compares the GPU performance for calculating absorption cross sections with different line profiles discussed by \citet{jt914} using \ExoMol\ $^{16}$O$^1$H MYTHOS dataset \citep{jt969}.

\begin{figure}
\centering
\begin{subfigure}{0.9\textwidth}
    \centering
    \includegraphics[width=\textwidth]{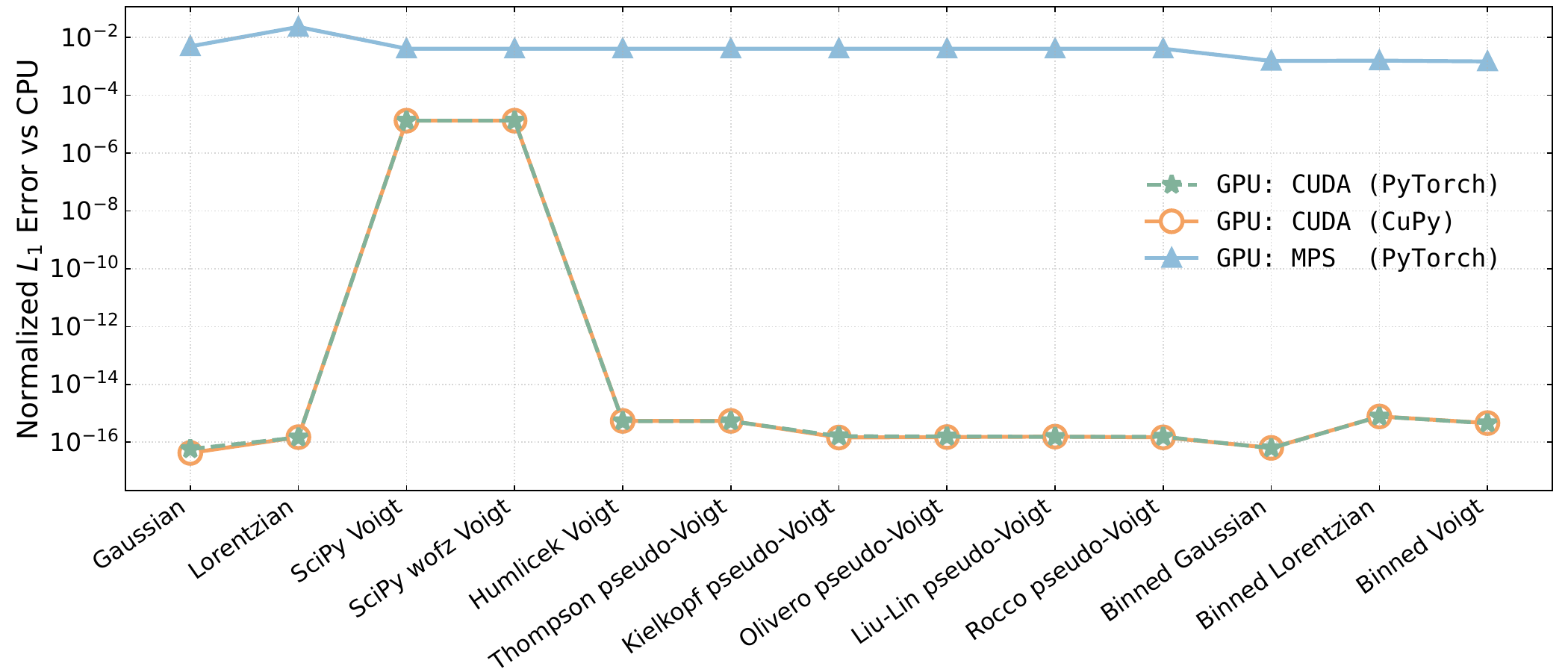}
    \caption{The Normalized $L_1$ error of each GPU backend}
\end{subfigure}
\begin{subfigure}{0.9\textwidth}
    \centering
    \includegraphics[width=\textwidth]{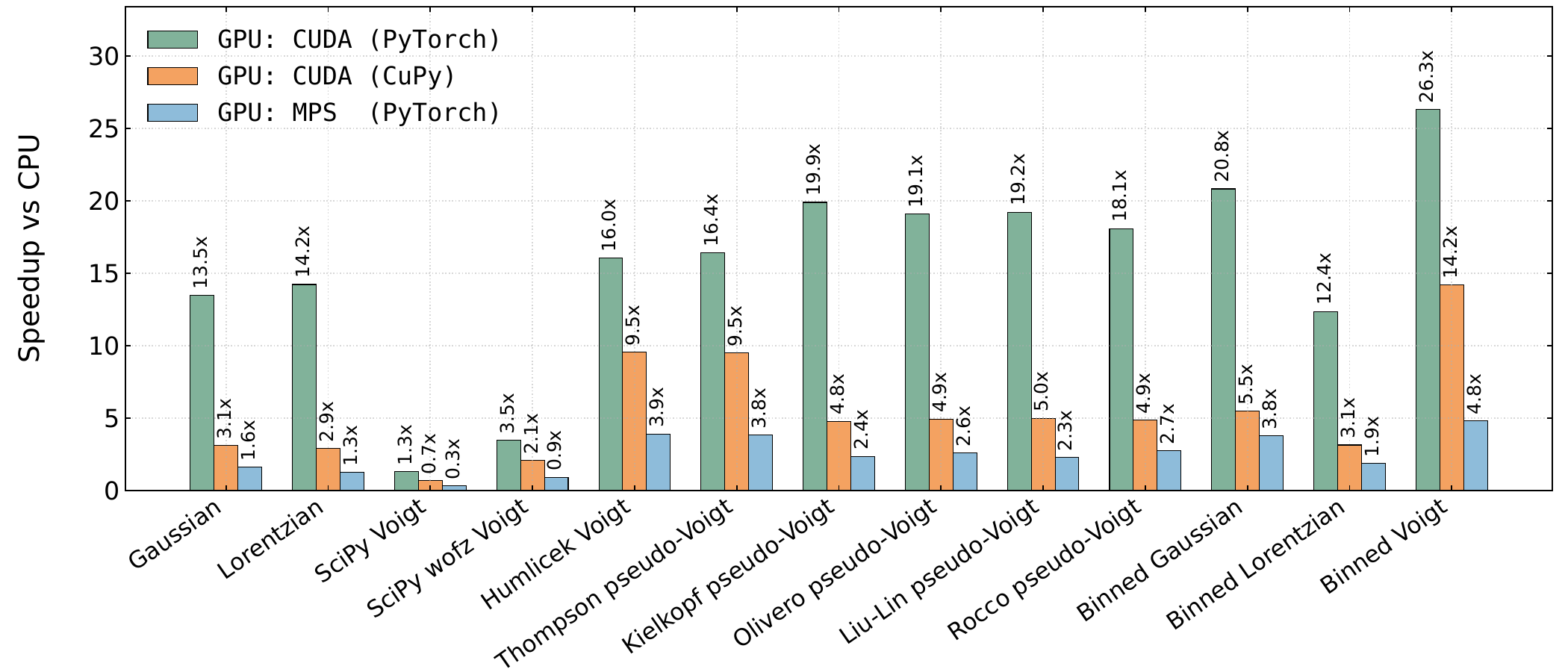}
    \caption{Speedup of each GPU backend}
\end{subfigure}
\begin{subfigure}{0.9\textwidth}
    \centering
    \includegraphics[width=\textwidth]{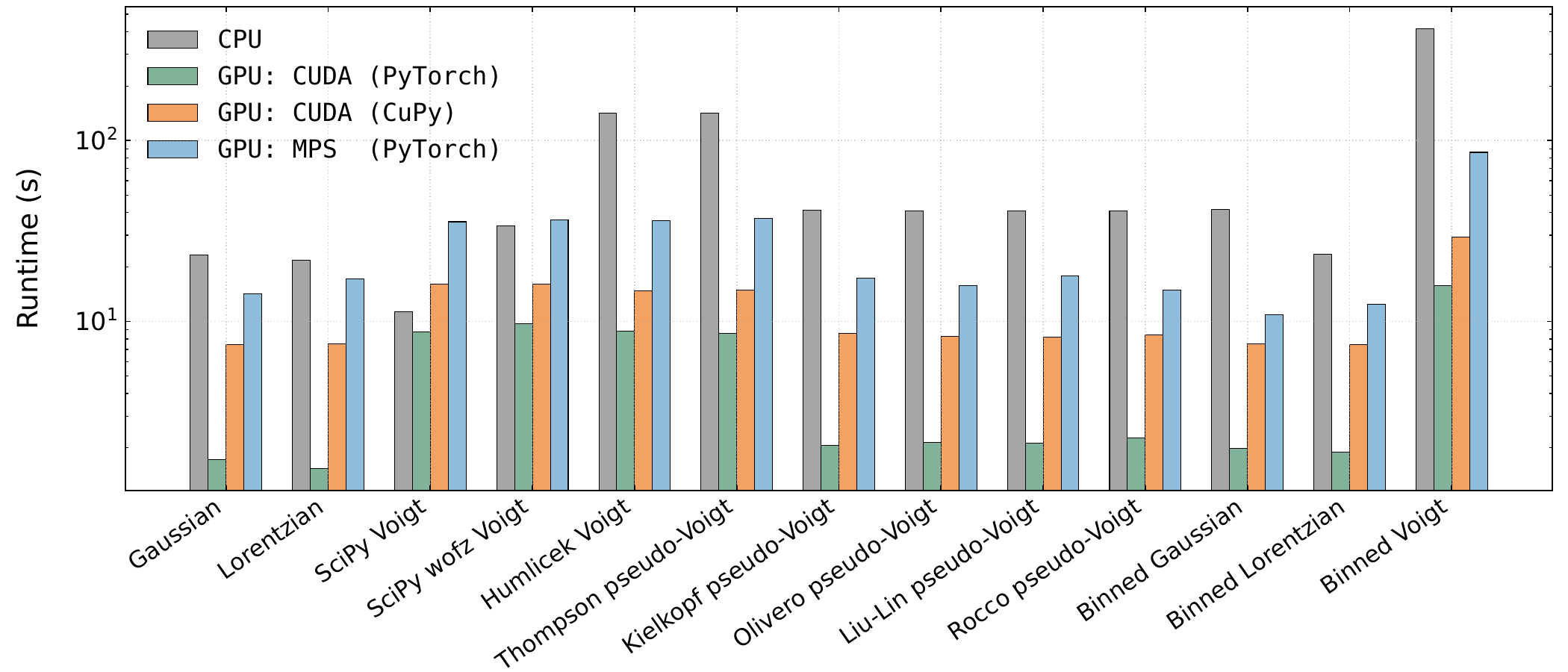}
    \caption{Absolute runtime: CPU vs GPU backends}
\end{subfigure}
\caption{Computational performance compared among CPU, GPU Torch provided CUDA (NVIDIA), GPU CuPy provided CUDA (NVIDIA), and GPU Torch provided MPS (MacOS) for calculating absorption cross sections with different line profiles discussed by \citet{jt914} using \ExoMol\ $^{16}$O$^1$H MYTHOS dataset.}
\label{fig:profiles_gpu}
\end{figure}

In Fig.~\ref{fig:profiles_gpu} (a), numerical agreement between each GPU backend and the CPU reference is quantified using the normalized $L_1$ error:
\begin{equation}
     \label{eq:profile_gpu_error}
    \epsilon_{L_1} = \frac{\sum_{i=1}^{N} \left| \sigma_{\rm GPU}(\tilde{\nu}_i) - \sigma_{\rm CPU}(\tilde{\nu}_i) \right| }{\sum_{i=1}^{N} \left| \sigma_{\rm CPU}(\tilde{\nu}_i) \right| }.
\end{equation}
This dimensionless metric measures the aggregate numerical discrepancy relative to the total magnitude of the CPU cross section. 
For most profiles, the CUDA backends agree with the CPU reference to approximately $10^{-16}$, close to double-precision numerical round-off. 
The normalized $L_1$ error increases to approximately $10^{-5}$ for SciPy Voigt and SciPy wofz Voigt because their CPU implementations use the SciPy Voigt or Faddeeva functions, whereas the corresponding GPU calculations use the Huml\'i\v{c}ek approximation when a native GPU Faddeeva function is unavailable.
These differences therefore arise primarily from the profile implementation. 
The MPS backend gives larger normalized $L_1$ errors  approximately in the range $10^{-3}$ to $2\times10^{-2}$, corresponding to aggregate discrepancies of about $0.1\%$--$2\%$. 
 These large MPS errors arise mainly from the use of
single-precision arithmetic float32 on MacOS's GPUs where
PyTorch MPS tensor operations are performed in single precision (\texttt{float32}), whereas the CPU and CUDA calculations use double precision (\texttt{float64}) in the present implementation. 
Additional small differences can arise from backend-specific special-function implementations and the order of batched floating-point summation.
The profile errors are evaluated using full-precision in-memory arrays, whereas the other errors in Fig.~\ref{fig:cf_stick_xsec_gpu} (a) are calculated from the saved output files. Consequently, zero error indicates agreement with the CPU reference at the numerical precision retained in the saved output files.

In Fig.~\ref{fig:profiles_gpu} (b), the speedup panel shows the ratio of runtime between GPU results and CPU result, see Eq.~(\ref{eq:gpu_speedup}).
\begin{equation}
    \label{eq:gpu_speedup}
    {\rm speedup}=\frac{t_{\rm CPU}}{t_{\rm GPU}}
\end{equation}
The largest gains occur for the most expensive profiles. 
On the RTX 4090, the PyTorch-CUDA backend reaches about $26.3\times$ for binned Voigt, $20.8\times$ for binned Gaussian, and $19.9\times$ for Kielkopf pseudo-Voigt. 
CuPy-CUDA is also faster than CPU for most profiles, but is generally much slower than PyTorch-CUDA in these tests. 
The exception is SciPy Voigt: CuPy-CUDA and MPS are slower than CPU for this profile because the GPU path has extra special function and data movement overhead relative to the relatively efficient CPU SciPy implementation.

In Fig.~\ref{fig:profiles_gpu} (c), the runtime panel confirms the same trend in absolute time. 
Simple Gaussian and Lorentzian profiles are already inexpensive on CPU, so the GPU benefit is modest. 
More complex Voigt and binned profiles are dominated by large tensor evaluations over the line and grid batches, where GPU parallelism is effective. 
Binned Voigt is the clearest case: the CPU runtime is about $415.58$ s, while PyTorch-CUDA reduces it to about $15.79$ s. 
MPS also accelerates several expensive profiles, but its benefit is smaller than CUDA because of lower hardware throughput and single precision tensor execution.

Overall, the new GPU mode is most useful for large line lists, wide wavenumber grids, and Voigt like or binned profiles. 
For production calculations requiring a good balance between speed and accuracy, PyTorch-CUDA is the preferred backend. 
Among the profiles tested, Huml\'i\v{c}ek Voigt is recommended when a Voigt-type profile is required, since it gives high speedup with negligible residuals. 
For pseudo-Voigt calculations, Kielkopf, Olivero, and Rocco pseudo-Voigt provide a good compromise between accuracy and runtime; 
Olivero pseudo-Voigt is a robust default choice. 
Binned Voigt gives the highest acceleration and is appropriate when bin integrated cross sections are desired, although its absolute runtime remains higher than the simpler pseudo-Voigt profiles. 
MPS is useful for local Apple silicon testing and moderate workloads, but CPU and CUDA should be used for final high throughput production runs.

\subsection{GPU performance for LTE and non-LTE} \label{sec:gpu lte nlte}

\begin{figure}
\centering
\begin{subfigure}{0.49\textwidth}
    \centering
    \includegraphics[width=\textwidth]{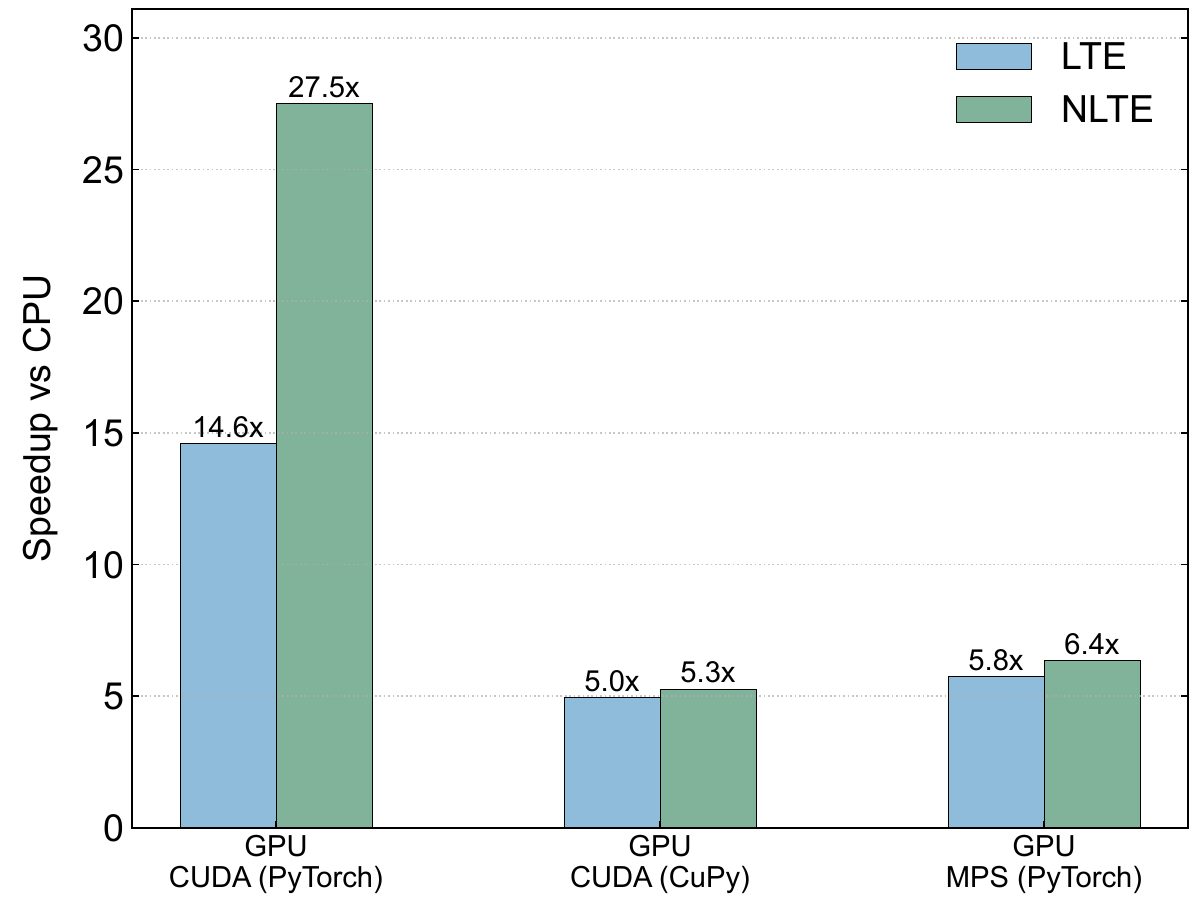}
    \caption{Speedup of each GPU backend}
\end{subfigure}
\begin{subfigure}{0.49\textwidth}
    \centering
    \includegraphics[width=\textwidth]{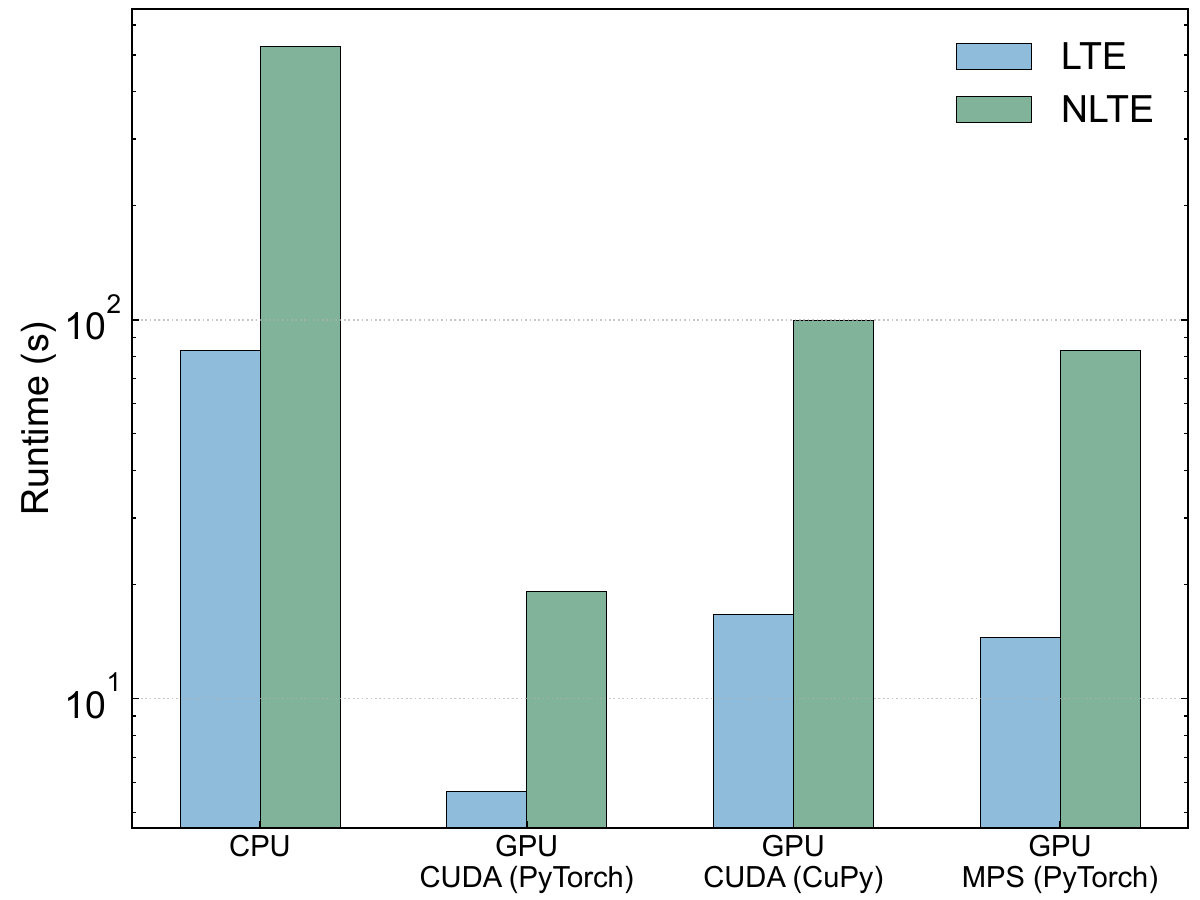}
    \caption{Absolute runtime: CPU vs GPU backends}
\end{subfigure}
\caption{Computational performance compared among CPU, GPU Torch provided CUDA (NVIDIA), GPU CuPy provided CUDA (NVIDIA), and GPU Torch provided MPS (MacOS) for calculating absorption cross sections with different LTE and non-LTE conditions using \ExoMol\ $^{16}$O$^1$H MYTHOS dataset.}
\label{fig:gpu_lte_nlte}
\end{figure}

GPU acceleration is particularly effective for two-temperature Treanor distributed non-LTE calculations than LTE calculations.  
Non-LTE does not merely rescale an LTE spectrum because it requires additional state population factors, altered stimulated emission terms, and repeated line strength evaluation over combinations of $T_{\rm rot}$ and $T_{\rm vib}$.  
These operations are dominated by large, independent array evaluations over transitions and spectral grid points, which map efficiently to GPU kernels.  
The CPU path pays more overhead for the extra population bookkeeping and repeated exponentials, whereas the GPU path amortises this cost through vectorised batches. 
Fig.~\ref{fig:gpu_lte_nlte} shows the speedup which is the ratio of runtime between GPU results and CPU result, see Eq.~(\ref{eq:gpu_speedup}) and runtime among difference modes.

\subsection{GPU accuracy and performance for cooling functions, stick spectra, and cross sections} \label{sec:gpu cf stick xsec}

This section compares the GPU performance for calculating cooling functions, LTE absorption stick spectra, and cross sections using the \ExoMol\ $^{16}$O$^1$H MYTHOS dataset. 
The CPU result is used as the reference in all cases. 
For a general output quantity $y$, the normalized $L_1$ error is defined as
\begin{equation}
    \epsilon_{L_1} = \frac{\sum_{i} \left| y_{\textrm{GPU},i} - y_{\textrm{CPU},i} \right| }{\sum_{i} \left| y_{\textrm{CPU},i} \right| }.
\end{equation}
where $y$ represents the cooling functions, intensities, or cross sections. 

\begin{figure}
\centering
\begin{subfigure}{0.9\textwidth}
    \centering
    \includegraphics[width=\textwidth]{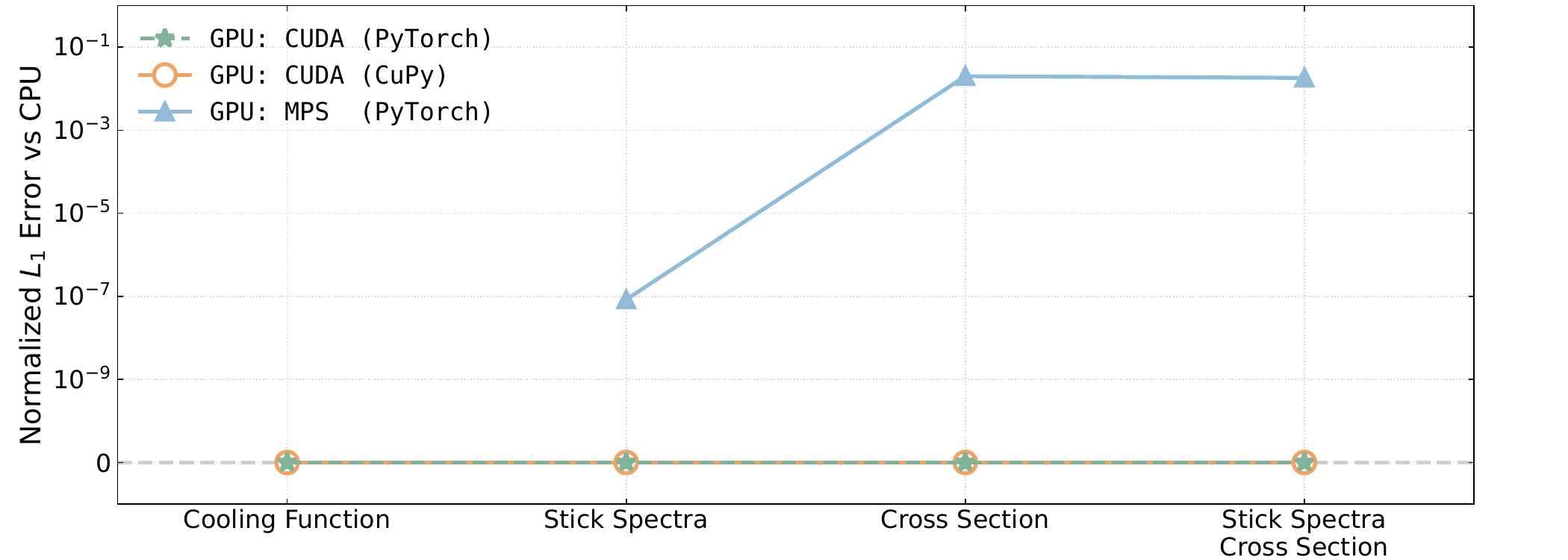}
    \caption{The normalized $L_1$ error of each GPU backend}
\end{subfigure}
\begin{subfigure}{0.9\textwidth}
    \centering
    \includegraphics[width=\textwidth]{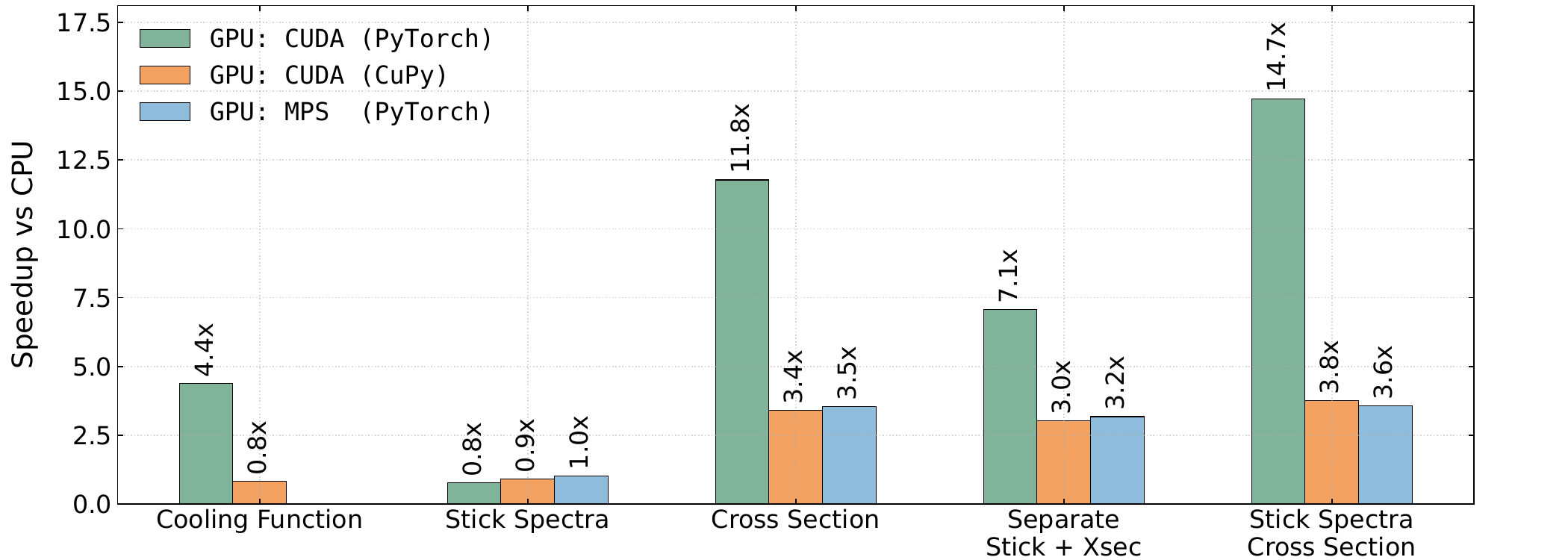}
    \caption{Speedup of each GPU backend}
\end{subfigure}
\begin{subfigure}{0.9\textwidth}
    \centering
    \includegraphics[width=\textwidth]{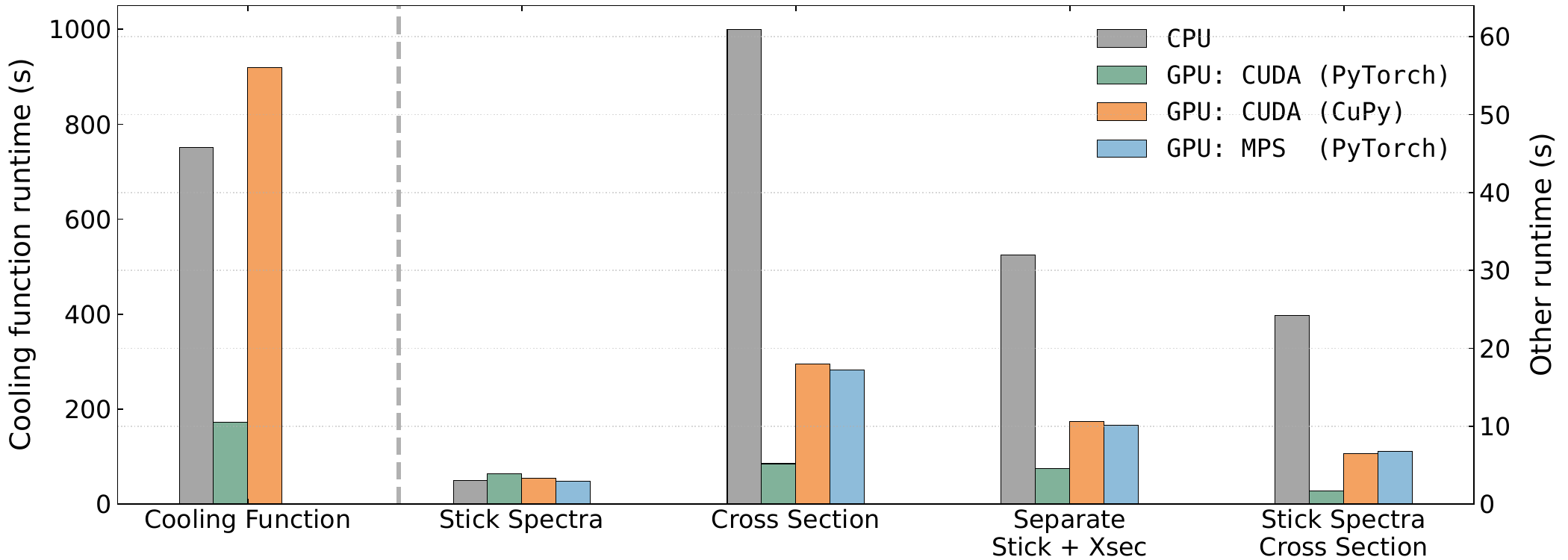}
    \caption{Absolute runtime: CPU vs GPU backends}
\end{subfigure}
\caption{Computational performance compared among CPU, GPU Torch provided CUDA (NVIDIA), GPU CuPy provided CUDA (NVIDIA), and GPU Torch provided MPS (MacOS) for cooling functions, LTE absorption stick spectra, and cross sections using the \ExoMol\ $^{16}$O$^1$H MYTHOS dataset. Zero error indicates agreement with the CPU reference at the numerical precision retained in the saved output files.}
\label{fig:cf_stick_xsec_gpu}
\end{figure}

In Fig.~\ref{fig:cf_stick_xsec_gpu} (a), the CUDA backends reproduce the CPU results at the precision of the saved files for all available LTE tests.
For the MPS backend, the normalized $L_1$ errors are approximately $8.36\times10^{-8}$ for stick spectra and $1.99\times10^{-2}$ for cross sections.
The latter corresponds to an aggregate discrepancy of approximately $1.99\%$ relative to the CPU cross section and primarily reflects the use of single-precision arithmetic by the PyTorch-MPS backend. 
No MPS result is shown for cooling functions because this calculation is not supported by the present PyTorch-MPS implementation. 
The cooling function calculation requires double-precision reductions over transitions and temperatures, while PyTorch-MPS provides incomplete support for \texttt{float64} operations. 
Consistent with this limitation, the MPS benchmark produced stick spectra and cross sections output files, but no cooling functions output. 
The MPS cooling functions entry is therefore left blank in Fig.~\ref{fig:cf_stick_xsec_gpu}.

In Fig.~\ref{fig:cf_stick_xsec_gpu} (b), the speedup panel shows that the strongest acceleration occurs for the LTE absorption cross sections calculation using Lorentzian line profile. 
The speedup is calculated using Eq.~(\ref{eq:gpu_speedup}).
The CPU runtime is about 130 s, while PyTorch-CUDA reduces it to about 19 s, giving a speedup of $6.9\times$. 
CuPy-CUDA and MPS also accelerate this calculation, with speedups of $2.9\times$ and $3.3\times$, respectively. 
This reflects the highly parallel structure of line profile evaluation over spectral grid points.

For cooling functions, PyTorch-CUDA reduces the runtime from about 751 s to 172 s, corresponding to a speedup of $4.4\times$. 
CuPy-CUDA is slower than CPU for this workload, with a speedup of $0.82\times$. 
This indicates that the cooling function calculation is more sensitive to backend implementation and memory movement than the cross sections calculation.

For stick spectra, the CUDA backends do not improve the runtime in this test. 
The CPU runtime is about 27 s, while CuPy-CUDA and PyTorch-CUDA take about 29 s and 30 s, respectively. 
The MPS backend is faster for this case, giving a speedup of $1.9\times$. 
The limited CUDA acceleration is expected because stick spectra timings depend on transition filtering, accumulation, and file output overheads, rather than dense spectral grid evaluation.

Fig.~\ref{fig:cf_stick_xsec_gpu} (c) shows the same trend in absolute runtime. 
GPU acceleration is most effective for cross sections, useful for cooling functions with PyTorch-CUDA, and limited for stick spectra. 
\PyExoCross\ can 

To summarise, PyTorch-CUDA gives the best performance among the tested backends while retaining CPU-level numerical agreement at the saved output precision.
For applications requiring high-accuracy cross sections, the CPU or CUDA backends are therefore preferred over MPS.

\subsection{Performance of the combined stick spectra and cross sections workflow for multiple temperatures and pressures} \label{sec:stick and xsec}

The runtime comparison in Fig.~\ref{fig:cf_stick_xsec_gpu} (c) also illustrates the advantage of the combined calculating stick spectra and cross sections workflow. 
\PyExoCross\ can now evaluate multiple temperatures and pressures in a single run, and in the \texttt{stick\_spectra\_cross\_section} mode the line list data are read and processed once before both quantities are computed. 
This reduces repeated I/O and preprocessing, including transition filtering and state matching, and is particularly beneficial for large line lists. 
The advantage is more pronounced when calculations are required over multiple thermodynamic conditions, since the same processed line information can be reused throughout the full workflow.
As shown in Fig.~\ref{fig:cf_stick_xsec_gpu} (c), the combined mode is faster than carrying out the stick spectra and cross sections calculations separately and summing their runtimes for all tested backends. 
For the CPU benchmark, the separate calculations require about 157 s in total, whereas the combined workflow takes about 127 s. 
The same behaviour is seen for the GPU backends.
Therefore, for production calculations involving large transition files and multiple thermodynamic conditions, the combined mode provides a more efficient execution pathway by amortising the data loading and preprocessing overhead across all requested outputs.

\section{Conclusions} \label{sec:sum}

\PyExoCross\ 2.0 is a significantly enhanced version of \PyExoCross\ \citep{jt914}; it uses molecular and atomic line lists to compute a range of spectroscopic and thermodynamic quantities, including partition functions, specific heats, cooling functions, radiative lifetimes, oscillator strengths, stick spectra, and cross sections. 
The new release supports two complementary interfaces: a standalone command-line Python program driven by input files, and a PyPI-distributed Python package with an API for use in scripts, notebooks, and automated workflows.

This version introduces optional GPU acceleration for selected large-scale calculations and extends the treatment of non-LTE stick spectra and cross sections through three methods: a two-temperature Treanor distribution, custom vibrational densities, and custom rotational populations. 
The required densities and populations are supplied through input files.

The computational workflow has also been improved. \PyExoCross\ can now evaluate multiple temperatures and pressures in a single run, and it includes a combined stick spectra and cross sections mode in which the line list data are read and processed once before both quantities are calculated. 
This reduces repeated I/O and preprocessing overhead, and improves efficiency for large line lists. 
In the present release, \PyExoCross\ supports these features exclusively for \ExoMol, \ExoMolHR, \ExoAtom, HITRAN, and HITEMP databases, with plans to expand functionality and support of additional data formats and databases for future versions.

\section*{Declaration of competing interest}

The authors declare that they have no known competing financial
interests or personal relationships that could have appeared to
influence the work reported in this paper.

\section*{Acknowledgements}

This work was supported by the European Research Council (ERC) under Advanced Investigator Project 883830 and by  STFC grants UKRI/ST/B001183  and ST/Y001508/1.

\section*{Data Availability}

The \PyExoCross\ program used for the calculations in this study is publicly available on GitHub at \href{https://github.com/ExoMol/PyExoCross}{https://github.com/ExoMol/PyExoCross}. Its Python package can be installed from PyPI via \href{https://pypi.org/project/pyexocross/}{https://pypi.org/project/pyexocross/}. The user documentation of \PyExoCross\ is accessible at \href{https://pyexocross.readthedocs.io/}{https://pyexocross.readthedocs.io/}.



\bibliographystyle{rasti}








\bsp	
\label{lastpage}
\end{document}